\documentclass[useAMS,usenatbib]{mn2e}
\usepackage[totalwidth=480pt, totalheight=680pt]{geometry}
\usepackage{graphics}
\usepackage{epsf}
\usepackage{subfigure}
\usepackage{amsmath}
\usepackage{amssymb}
\def\ni{\noindent}

\begin{document}

\title[Non-Linear LMC PL Relations]{Further empirical evidence for the non-linearity of \\
the period-luminosity relations as seen in the Large \\
Magellanic Cloud Cepheids}
\author[Ngeow et al.]{Chow-Choong Ngeow$^{1,5}$\thanks{E-mail: cngeow@astro.uiuc.edu}, Shashi M. Kanbur$^{1,6}$, Sergei Nikolaev$^{2}$, John Buonaccorsi$^{3}$,\newauthor 
Kem H. Cook$^{2}$ and Douglas L. Welch$^{4}$
\\
$^{1}$Department of Astronomy, University of Massachusetts,         
 Amherst, MA 01003, USA
\\
$^{2}$Institute for Geophysics and Planetary Physics, Lawrence Livermore National Laboratory, Livermore, CA 94550, USA
\\
$^{3}$Department of Mathematics and Statistics, University of Massachusetts,         
 Amherst, MA 01003, USA
\\
$^{4}$Department of Physics and Astronomy, McMaster University, Hamilton, ON L85 4M1, Canada
\\
$^{5}$Department of Astronomy, University of Illinois, Urbana-Champaign, IL 61801, USA
\\
$^{6}$Department of Physics, State University of New York at Oswego, Oswego, NY 13126, USA
}

\date{Accepted 2005 month day. Received 2005 month day; in original form 2005 April. 11}
\maketitle

\begin{abstract}

    Recent studies, using OGLE data for LMC Cepheids in the optical, strongly suggest that the period-luminosity (PL) relation for the Large Magellanic Cloud (LMC) Cepheids shows a break or non-linearity at a period of 10 days. In this paper we apply statistical tests, the chi-square test and the $F$-test, to the Cepheid data from the MACHO project to test for a non-linearity of the {\it V}- and {\it R}-band PL relations at 10 days, and extend these tests to the near infrared ({\it JHK}-band) PL relations with 2MASS data. We correct the extinction for these data by applying an extinction map towards the LMC. The statistical test we use, the $F$-test, {\it is} able to take account of small numbers of data points and the nature of that data on either side of the period cut at 10 days. With our data, the results we obtained imply that the {\it VRJH}-band PL relations are non-linear around a period of 10 days, while the {\it K}-band PL relation is (marginally) consistent with a single-line regression. The choice of a period of 10 days, around which this non-linearity occurs, is consistent with the results obtained when this "break" period is estimated from the data.

    We show that robust parametric (including least squares, least absolute deviation, robust regression) and non-parametric regression methods, which restrict the influence of outliers, produce similar results. Long period Cepheids are supplemented from the literature to increase our sample size. The photometry of these long period Cepheids is compared with our data and no trend with period is found. Our main results remain unchanged when we supplement our dataset with these long period Cepheids. By examining our data at maximum light, we also suggest arguments why errors in reddening are unlikely to be responsible for our results. The non-linearity of the mean {\it V}-band PL relation as seen in both of the OGLE and MACHO data, using different extinction maps, suggests that this non-linearity is real.

\end{abstract}

\begin{keywords}
Cepheids -- Stars: fundamental parameters
\end{keywords}

\section{Introduction}

     The Cepheid period-luminosity (PL) relation plays a crucial role in the distance ladder which can ultimately be used to determine Hubble constant. The current most widely used PL relation is based on the Large Magellanic Cloud (LMC) Cepheids \citep[e.g., see][]{fre01,sah01}. The Cepheid PL relation has long been considered to be a linear function of $\log(P)$ within the range of $\log(P)\sim0.3$ to $\log(P)\sim2.0$, where $P$ is the pulsation period in days. Current versions of the LMC PL relations can be found, e.g., in \citet{mad91}, \citet{tan97}, \citet{gie98}, \citet{uda99} and \citet{per04}.

     However, recent studies \citep{tam02,tam02a,kan04,san04} strongly suggest that the LMC PL relation is not linear, as there are two PL relations for the short ($P<10$ days) and the long ($P>10$ days) period LMC Cepheids, respectively. These studies are mainly based on the optical data ({\it BVI}-band) from the Optical Gravitational Lensing Experiment \citep[OGLE,][]{uda99,uda99b} database for the fundamental mode Cepheids.  The non-linear {\it V}- and {\it I}-band LMC PL relations are further supported by the results from a rigorous statistical test, the $F$-test, as described in \citet[][hereafter KN]{kan04}. This $F$-test {\it is} sensitive to the number of data points and the nature of the data on either side of 10 days. This sensitivity is such that if, either the number of data points is small or the data has a large scatter, the slopes on either side of 10 days are not well defined and the test returns a non-significant result. We describe this property of the $F$-test in detail in Section 3. Hence the work of KN, which used the OGLE data alone and found marked non-linearities in the PL (and PC, period-colour) relations in the LMC at a period of 10 days is in contrast to \cite{gie05}, who state that the OGLE data {\it alone} do not show any non-linear behavior or break at a period of 10 days.
Since the OGLE Cepheid data are truncated at $\log(P)\sim1.5$ \citep{uda99}, \citet{san04} used additional {\it BVI}-band data that are available
from the literature, especially those with $\log(P)>1.5$, to further support the existence of two PL relations in the LMC Cepheids. Compared to previous studies, the large number of LMC Cepheids with high quality photometric data, together with the estimation of extinction ($E[B-V]$) to individual Cepheids,
given in the OGLE database has made the detection of non-linear LMC PL relation becomes possible.

     The non-linearity of Cepheid PL relations has been recognized for some time in the literature, as the mean magnitudes for some of the long period Cepheids ($\log[P]\gtrsim 1.5$) do not follow the best-fit lines. In fact, some of the earlier studies (including the works of Kukarkin, Shapley \& Fernie) on the Cepheid PL relation have suggested or used a quadratic form, $M=\alpha+\beta \log(P)+\gamma [\log(P)]^2$, to describe the PL relation \citep{fer69}. For example, a semi-empirical derivation of the quadratic PL relations in the {\it B}- and {\it V}-band can be found in \citet{fer67}. In addition, the composite PL relations at mean and maximum light constructed with Cepheids in different galaxies, as presented in \citet{san68}, also show some curvature of the PL relations. In terms of theoretical modeling, \citet{bon99}, \citet{cap00} and \citet{mar05} have fit quadratic PL relations to the theoretical periods and magnitudes obtained from the pulsational modeling. The non-linearity of the PL relations, however, can also be explained with the broken PL relations: there are two PL relations, for short and long period Cepheids respectively, with a break at a specific period (e.g., at 10 days). Perhaps the earliest idea for the broken PL relation was proposed by Kukarkin in 1937, as quoted from \citet{fer69}: ``...he concluded that the relation was definitely non-linear...but the best fit was by two straight lines of different slope intersecting at a period of 10 days''. The present-day version of the broken PL relations is presented in \citet{tam02}, KN and \citet{san04} with the OGLE data, as mentioned in the previous paragraph. These broken PL relation are also compared with the theoretical predictions in \citet{mar05}, which show a quite satisfactory agreement. Note that \citet{cap99} have fit a broken PL relation to their results obtained from the theoretical models, but with a break at $\log(P)=1.4$. 

     Besides the non-linear PL relations seen in the LMC Cepheids, the SMC fundamental mode Cepheids also show a change of slope for Cepheids with periods less than 2 days from the EROS-2 data \citep{bau99}. The authors carefully examine the possible errors or bias in the data that lead to this finding, and conclude that the non-linear PL relations seen in SMC Cepheids is real. \citet{bar98} explained that this change of the slope for short period (less than 2 days) SMC Cepheids is due to evolutionary effects: this is because the blue loops for these short period SMC Cepheids do not fully extend into the instability strip. However, the non-linearity of the LMC PL relation at $\sim10$ days period is believed to be independent of this evolutionary effect. In this paper, we concentrate on studying the 10 days non-linearity for the LMC PL relation and do not suggest that this is the same phenomena described by \cite{bar98} but for the higher metallicity LMC Cepheids.

     The motivation of this paper is to extend the work of KN into the study of the linear and/or non-linear nature of the PL relations at/around 10 days for the fundamental mode LMC Cepheids. It it important to verify the non-linearity of the LMC PL relation at 10 days from the studies of \citet{tam02}, KN and \citet{san04}, which mainly analyzed the OGLE data, with an independent dataset. Hence, we examine the non-linearity of the optical LMC PL relation with the Cepheid data obtained from the MACHO database \citep{alc00} in this paper. In addition, we also investigate the LMC PL relations in the near infrared (NIR) bands (i.e., {\it JHK}-band) to see if the NIR PL relations also show a non-linearity. We apply statistical tests to investigate the possible non-linearity of the LMC PL relation in these bands. We formulate this non-linearity in three ways and examine if the data are more consistent with a linear or non-linear PL relation.

     In this paper, we only use the published mean magnitudes (and periods) for the LMC Cepheids from the literature. The data used in this paper is described in Section 2. The statistical analysis and the results are presented in Section 3 and 4, respectively. The conclusion \& discussion will be followed in Section 5. 

\section{Data and Extinction Corrections}

     \begin{figure*}
       \vspace{0cm}
       \hbox{\hspace{0.2cm}\epsfxsize=8.5cm \epsfbox{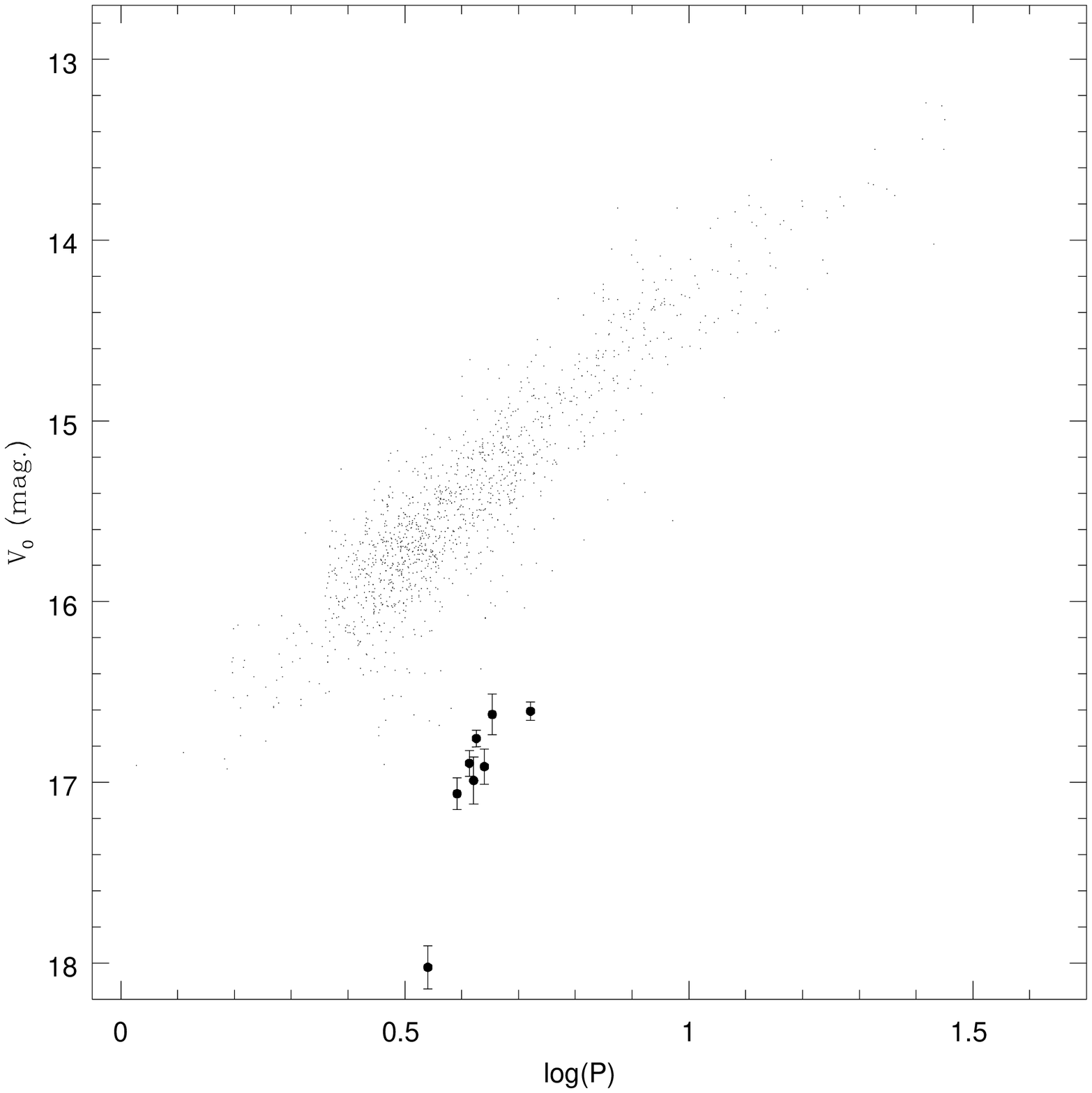}
         \epsfxsize=8.5cm \epsfbox{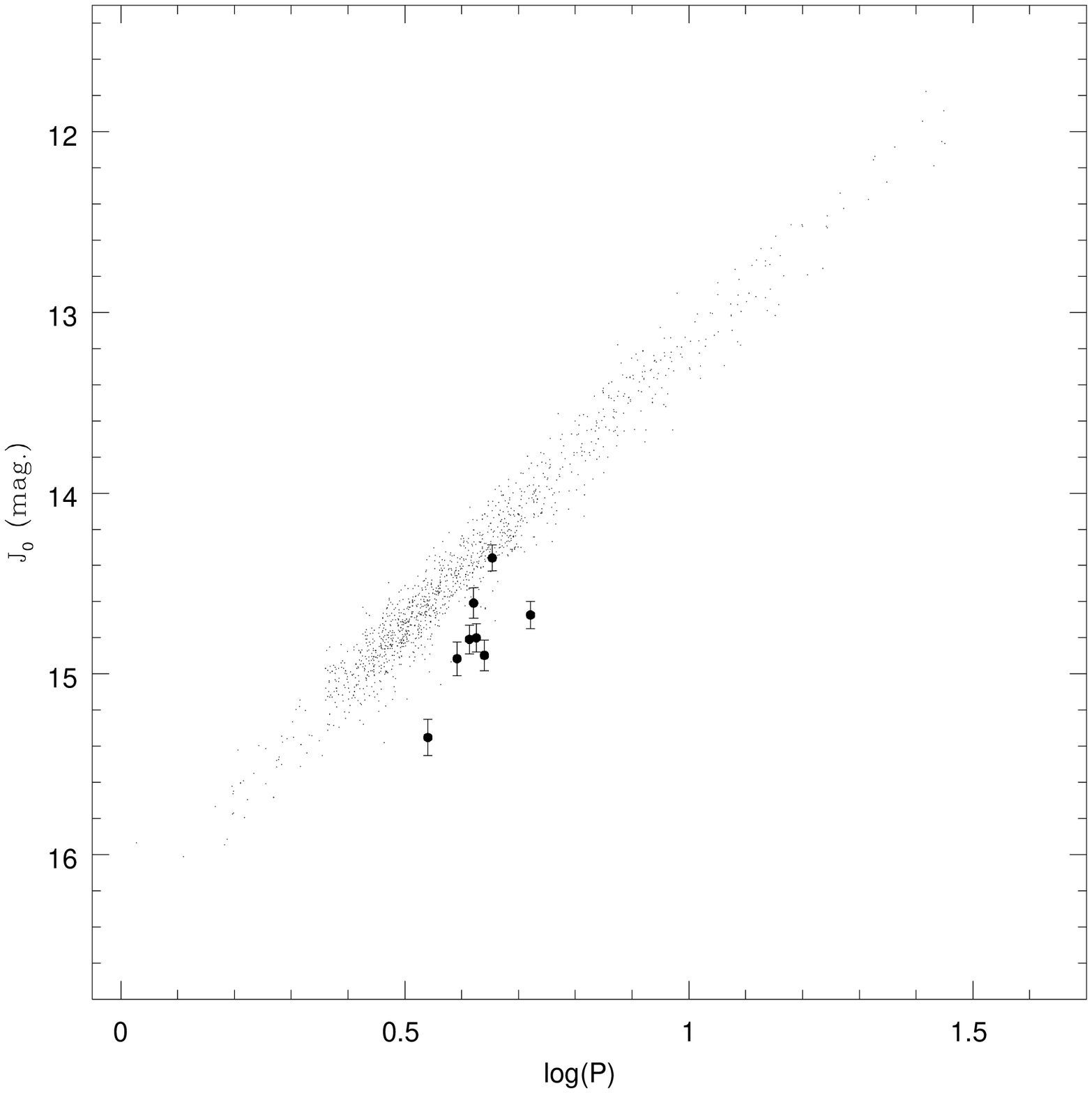}}
       \vspace{0cm}
       \caption{The extinction corrected {\it V}- and {\it J}-band PL relations with all 1330 Cepheids in the sample. The filled circles are the 8 outliers as described in the text. The error bars for the outliers are the quadrature sum of the photometric errors and the extinction errors. The error bars for other Cepheids are omitted for clarity.}
       \label{figoutlier}
     \end{figure*}

      The data for the 1330 fundamental mode LMC Cepheids are adopted from \citet{nik04}. This includes the {\it V}- and {\it R}-band mean magnitudes, as well as the periods, from the MACHO database and the {\it J}-, {\it H}- and {\it K}-band mean magnitudes from the Two Micron All Sky Survey database \citep[2MASS,][]{skr98}. The mean magnitudes in the {\it V}- and {\it R}-bands are computed from the complete light curves with the MACHO photometric data. The (random phase) magnitudes in {\it J}-, {\it H}- and {\it K}-bands are from the 2MASS single epoch observations, hence the mean magnitudes in these NIR bands are obtained with a random-phase correction function. The details for obtaining the mean magnitudes in all these bands can be found in \citet{nik04}. To check the accuracy of these {\it JHK} mean magnitudes, one of us (Nikolaev et al, 2005 -- in preparation) has compared the {\it JHK} mean magnitudes from the 2MASS data (after random-phase correction) with the \citet{per04} data, which are obtained from the well-sampled ({\it JHK}) light curves. For the 33 matched Cepheids in the MACHO/2MASS sample and the \citet{per04} sample, the average difference in the mean magnitudes are: $\overline{\Delta_J}=-0.022$ (with std $=0.070$, where std $=$ standard deviation), $\overline{\Delta_H}=-0.012$ (std $=0.073$) and $\overline{\Delta_K}=-0.015$ (std $=0.068$). These are completely consistent with the zero difference. More importantly, the plots of the difference between the means in these bands as a function of period (not shown) do not show any trends with period, hence the {\it JHK} mean magnitudes from the 2MASS database will not seriously affect our analysis and results. We did not use other datasets in {\it VRJHK}-bands, as in \citet{san04}, to perform the main analysis in order to keep our dataset as consistent and homogeneous as possible (however, see Section 5.2). 

      The next step in the analysis is to obtain the extinction-corrected mean magnitudes for these Cepheids. This was done by applying an extinction map with the following values of $R_{\lambda}$ (ratio of total-to-selective reddening, and $R_{\lambda}=A_{\lambda}/E[B-V]$), in order to be consistent with \citet{nik04}: $R_{V,R,J,H,K} = \{3.12,\ 2.56,\ 0.90,\ 0.53,\ 0.34\}$. The LMC extinction map we used is taken from \citet{zar04}. In brief, given the coordinates for each Cepheid and a search radius $r$, the extinction map returns the average value of extinction, $\overline{A_V}\pm\sigma_{A\mathrm{v}}$ (where $\sigma_{A\mathrm{v}}=\mathrm{std}/\sqrt{N}$), from the stars bounded within the search radius. The extinction map allows us to choose the $\overline{A_V}$ obtained from cool stars ($5500\mathrm{K}<T_{eff}<6500\mathrm{K}$), hot stars ($12,000\mathrm{K}<T_{eff}<45,000\mathrm{K}$) or both. Since Cepheid variables are cool stars, we only use the extinction values from the cool stars. We use $r=2'$ to ensure that there are sufficient numbers of stars within the search radius to obtain the $\overline{A_V}$ (usually more than 10 stars)\footnote{Only two Cepheids with enclosed stars that are less than 10}. However, there are two Cepheids with locations beyond the extinction map and one Cepheid without any cool stars within the $r=3'$ radius. For simplicity, we assume $\overline{A_V}=0.3\pm0.1$ mag (roughly the mean extinction of LMC) for these three Cepheids. After the values of $\overline{A_V}$ for the Cepheids are obtained, the extinctions in other bands are calculated with the following relation: $A_{\lambda}=(\frac{R_{\lambda}}{R_V})\overline{A_V}$ with the errors $\sigma_{A_{\lambda}}=(\frac{R_{\lambda}}{R_V})\sigma_{A\mathrm{v}}$, where $\lambda$ denotes the passband.      

      We consider the following three samples in this study:
      \begin{description}
      \item {\tt Sample A} - This sample consist of all 1330 Cepheids in our dataset.
      \item {\tt Sample B} - We then remove 8 outliers out of 1330 Cepheids (all of these outliers have $\log[P]<0.75$) from Sample A, as they are fainter by more than $\sim1.2$mag. and $\sim1.0$mag. in {\it V}- and {\it R}-band, respectively, to the fitted PL relations at the same period. The locations of these outliers in the PL relations are plotted in Figure \ref{figoutlier}. It is possible that the extinctions for these outliers are underestimated because they are not obvious in the NIR PL relations (see the right panel of Figure \ref{figoutlier}). Another possibility is that they are Type II Cepheids that are mis-identified as the classical Type I Cepheids. However, it is clear that these outliers should be removed from our sample. Hence Sample B consists of 1322 Cepheids.
      \item {\tt Sample C} - Following \citet{uda99}, we also remove the Cepheids with $\log(P)<0.4$ from Sample B. This is to guard against the possible contamination from first overtone Cepheids \citep{uda99}, as well as other types of variables such as the anomalous Cepheids and the double-mode Cepheids \citep{alc99}. In addition, this choice of the period cut will also guard against the possible change of slopes for short ($P<2$ days) period Cepheids, as seen in the SMC Cepheids \citep{bau99,uda99}. Note that \citet{uda99} has applied the period cut of $\log(P)=0.4$ to the OGLE LMC Cepheids to derive the PL relations that were used in other studies, such as the $H_0$ Key project \citep{fre01}. Hence the same period cut was also applied in previous studies of the non-linear LMC PL relations \citep{tam02,kan04,san04}. Furthermore, the period distribution of the LMC Cepheids shows a sharp break at $\log(P)\sim0.4$ \citep{alc99}. This is probably due to evolutionary effects \citep{bar98,alc99}. We contend that these reasons justify the removal of Cepheids with $\log(P)<0.4$, and leaves 1216 Cepheids in this sample. Our use of the $F$-test, as given in the next section, ensures that such a period cut will not affect our results. 
      \end{description}

\section{Analysis}

      In order to study the linear and/or non-linear nature of the LMC PL relation, the data presented in the previous section are fitted with the (weighted) least squares regression \citep[see, e.g.,][]{pre92}. Ordinary least squares regression, $OLS(Y|X)$, has been extensively used in the literature to obtain the fitted PL relations from observations \citep[for examples, see][and many other papers on PL relation using the Galactic Cepheids]{mad91,lan94,gie98,uda99,kan04,san04,fre01} as well as from theoretical modelings \citep[as in][]{bon99,cap00,bar01}. This is because there is a good physical theory \citep[see, e.g.,][]{san58,cox80,mad91} suggesting why the period is the independent variable and is measured essentially without errors, hence the $OLS(Y|X)$ regression can be applied. In terms of observations, the errors of the periods for the MACHO Cepheids used in this paper (see Section 2) is about $10^{-4}P$ \citep{nik04}, which is comparable to the accuracy reported by the OGLE team \citep[$\sigma_P\sim10^{-5}P$,][]{uda99b}, as well as other high quality photometric observations (e.g., see \citealt{mof98}, and some examples from the Galactic Cepheids in \citealt{cou85}, \citealt{sho92} and \citealt{ber94}). The $OLS(Y|X)$ regression is also recommended for the distance scale studies \citep{iso90,fei92}.
       
     The weights ($w$) that are used in the regressions are the quadrature sum of the photometric errors in mean magnitudes ($\sigma_{\lambda}$), the errors from the random phase corrections that are only applicable to {\it JHK}-bands ($\sigma_{rp}$), the errors from extinction ($\sigma_{A_{\lambda}}$), and the intrinsic dispersion due to the finite width of the instability strip ($\sigma_{IS}$). Hence, $w(i)=\sigma^2_{\lambda}(i)+\sigma^2_{rp}(i)+\sigma^2_{A_{\lambda}}(i)+\sigma^2_{IS}$. The dispersions of the PL relations in each band from \citet{mad91} can be initially adopted for $\sigma_{IS}$. However, a comparison of the dispersions given in \citet{mad91} with \citet{gie98}, \citet{per04} and \citet{san04} suggest that the {\it VR}-band and the {\it JHK}-band dispersions given in \citet{mad91} should be scaled by $\sim0.73$ and $\sim0.88$, respectively. Hence we adopt the following values for the intrinsic dispersions: $\sigma_{IS}(V,R,J,H,K) = \{0.21,\ 0.18,\ 0.14,\ 0.12,\ 0.11\}$. The adopted $\sigma_{IS}$, such as those originally given in \citet{mad91}, does not affect the results from the following statistical tests. We also assume that all LMC Cepheids are at the same distance so that there is no distance variation in the weight formula.
    
\begin{figure}
       \hbox{\hspace{0.1cm}\epsfxsize=7.5cm \epsfbox{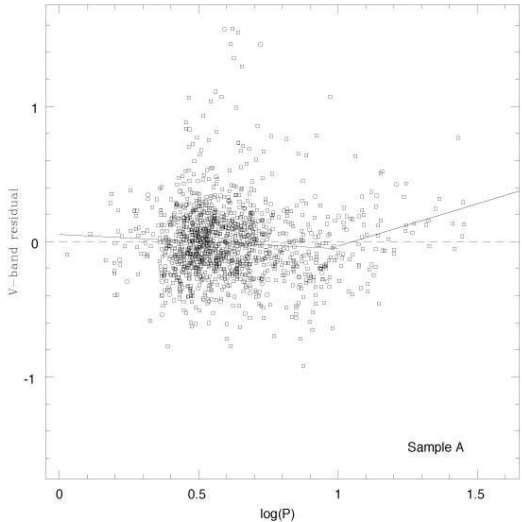}}
       \caption{The residuals for the {\it V}-band PL relation as a function of log-period from Sample A. The residuals are obtained assuming the one-line regression is adequate. The solid lines are the linear fits of the residuals as function of log-period, separated at 10 days. The dashed line is the zeroth residual line. The residual plots from Sample B \& C are similar to this plot.}
      \label{figres}
     \end{figure}

     One way to check for a non-linearity on the LMC PL relation is through a residual plot. If the one-line regression is adequate, then the residuals should be populated (roughly) evenly around the zeroth residual line (i.e., independent of period). On the other hand, if the residuals show a trend, or some trends, in the residual plot, then more parameters are needed to fit the data. The residuals of the MACHO {\it V}-band data from previous section are plotted as function of $\log(P)$ in Figure \ref{figres}, assuming that the one-line regression is adequate. A linear fit of the residuals with $\log(P)>1.0$ gives non-zero slopes of $-0.66\pm0.27$, $-0.65\pm0.27$ and $-0.67\pm0.27$ from Sample A, B and C, respectively. These slopes clearly indicate that the trends seen in Figure \ref{figres} are significant. Hence a one-line regression may not be adequate to fit the MACHO {\it V}-band PL relations. Similar residual plots in other bands are presented in the right panels of Figure \ref{fig1}.

     To investigate the possible non-linearity of the PL relations, we applied the chi-square test and the $F$-test \citep[as given in][and used initially in KN]{wei80} using the weighted least squares results. The null ($H_O$) and the alternate ($H_A$) hypothesis represent the reduced and the full models, respectively, where the reduced models are obtained by setting some parameters in the full models to certain specific values (e.g., zero, see \citealt{wei80}). Hence our reduced model is a one-line regression (2-parameter) fit to the entire period range, and the null hypothesis, $H_O$, is that one-line regression is adequate to describe the data. In contrast, $H_A$ is that the full model is necessary to fit the data. Here, we consider three different kinds of full models: 

     \begin{enumerate}
     \item A two-line regression (4-parameter) with a break at a characteristic period, $P_o$, i.e.,
       \begin{eqnarray}
         m &=& a^L\log(P) + b^L, P > P_o, \\ 
         m &=& a^S\log(P) + b^S, P < P_o. \nonumber
       \end{eqnarray}

     \item A piecewise continuous regression (3-parameter) that forces the two-line regression to be continuous at $P_o$, i.e.,

       \begin{eqnarray}
         m &=& c + d\times \mathrm{min}[\log(P),\log(P_o)] \nonumber \\
           & & + e\times \mathrm{max}[0,\log(P)-\log(P_o)]. 
       \end{eqnarray}

     \item A quadratic regression (3-parameter) of the form of:

       \begin{eqnarray}
         m &=& \alpha + \beta\log(P) + \gamma [\log(P)]^2.
       \end{eqnarray}
         
     \end{enumerate}

     \ni Although we consider these three models, our main interest is in the investigation of a change of slope in the LMC PL relation for short and long period Cepheids. It is clear that two lines of differing slope, continuous at $P_o$ and even perhaps with a small discontinuity at $P_o$, are approximately a quadratic with a local minimum/maximum derivative at $P_o$.

     To apply the chi-square test and the $F$-test, we then calculated the weighted error sum of squares, $SSE = \sum [m(i) - \hat{m}(i)]^2/w(i)$, for both $H_O$ and $H_A$, where $m(i)$ and $\hat{m}(i)$ are the observed magnitudes and the fitted magnitudes from weighted least squares for $i{\mathrm th}$ Cepheid, respectively. Under $H_O$, the difference of the $SSE$: 
     
     \begin{eqnarray}
       \chi^2 = SSE(H_O) - SSE(H_A), 
     \end{eqnarray}
       
     \ni follows approximately a chi-square distribution with $\nu=n_{FL}-n_{RD}$ degrees of freedom \citep[e.g, see][]{wei80}, where $n_{FL}$ \& $n_{RD}$ are the number of parameters in full and reduced models, respectively. For a large number of data points ($N$), this is essentially the same as applying a standard $F$-test (which many regression packages produce) under weighted least squares, where 
     
     \begin{eqnarray}
       F = \frac{\chi^2/\nu}{SSE(H_A)/(N-n_{FL})}, 
     \end{eqnarray}

     \ni which is approximately a $F_{(\nu,N-n_{FL})}$ distribution under the null hypothesis \citep[see also][]{kan04}, where $\chi^2$ is given in equation (4). The corresponding probability ($p$) for the $\chi^2$ and the $F$-values, under $H_O$, can be obtained from consulting the standard statistical books or softwares. A large value of $\chi^2$ and/or $F$ (hence small $p$) indicates that the null hypothesis can be rejected, and therefore the data is more consistent with the alternate hypothesis. Table \ref{tabp} summarizes the $\chi^2$ and $F$-values that correspond to the case of $p=0.05$ and $p=0.01$. In case of the $F$-test, $p(F)<10^{-3}$ for $F>7$ with large number of data points (say, $N>500$).

    The $F$-test relies on the assumptions that the residuals are normally distributed and are homoskedastic. KN tested the OGLE data for these assumptions. With such large numbers of data departures from normality are unlikely to significantly influence our results. The chi-square test, as formulated above, does not depend on homoskedasticity and hence we use it in this study to supplement our $F$-test results. Perhaps a simpler chi-square test is the following: compare the chi-square statistic for a one- and two-line regression respectively. However, this approach does not have the necessary power to distinguish between the two cases and hence we use the chi-square test given in equation (4) above.
     
     \begin{table}
       \centering
       \caption{The critical values for the chi-square test and the $F$-test.}
       \label{tabp}
       \begin{tabular}{lcc} \hline 
         $\nu^{\mathrm a}$  & $p=0.05$ & $p=0.01$ \\     
         \hline 
         \multicolumn{3}{c}{$\chi^2$ values} \\ 
         1  & 3.84 & 6.63 \\
         2  & 5.99 & 9.21 \\
         \multicolumn{3}{c}{$F$ values} \\ 
         1  & 3.84 & 6.63 \\
         2  & 3.00 & 4.61 \\
         \hline
       \end{tabular}
       \begin{list}{}{}
       \item   $^{\mathrm a}$ $\nu=1$ for the alternate hypothesis given in equation (2) \& (3); $\nu=2$ for the alternate hypothesis given in equation (1). 
       \end{list} 
     \end{table}

     \begin{table*}
       \centering
       \caption{Testing the significance of the $F$-test with various period cuts.}
       \label{tabpcut}
       \begin{tabular}{cccccccccc} \hline 
                      &             & \multicolumn{4}{c}{OGLE} & \multicolumn{4}{c}{MACHO} \\
         $\log(P_1)$  & $\log(P_2)$ & $N_{\log P<1}$ & $N_{\log P>1}$ & $F$ & $p(F)$ & $N_{\log P<1}$ & $N_{\log P>1}$ & $F$  & $p(F)$ \\     
         \hline 
         0.4 & 1.5 & 585 & 49 & 8.867 & 0.000  & 1147 & 69 & 12.32 & 0.000 \\
         0.6 & 1.4 & 268 & 45 & 8.799 & 0.000  & 494  & 63 & 10.62 & 0.000 \\
         0.7 & 1.3 & 136 & 41 & 6.904 & 0.001  & 249  & 58 & 5.438 & 0.005 \\
         0.8 & 1.2 & 72  & 30 & 0.279 & 0.757  & 141  & 51 & 1.124 & 0.327 \\
         0.9 & 1.1 & 25  & 15 & 0.909 & 0.412  &  56  & 29 & 0.029 & 0.972 \\
         0.9 & 1.5 & 25  & 49 & 1.000 & 0.373  &  56  & 69 & 0.225 & 0.799 \\
         \hline
       \end{tabular}
     \end{table*}

     Furthermore, the $F$-test we use, as in equation (5), to test for changes in slope on either side of a period cut, is sensitive to the number and nature of Cepheid data (such as the distribution of the data) defining the slope on either side of the period cut. Equation (5) contains an explicit dependence on these factors. Since the variance of the fitted slope, $\hat{a}$, is $\mathrm{Var}(\hat{a})=\frac{1}{(N-n)}\frac{SSE}{S_{XX}}$, where $n$ is number of parameters and $S_{XX}=\sum(x-{\overline x})^2$ (${\overline x}$ is the mean value of $x=\log[P]$), hence another way of writing $SSE$ as used in equation (5) is $SSE=\mathrm{Var}(\hat{a})(N-n)S_{XX}$. Thus the $F$-test automatically builds into it the amount of precision in both slopes, which in turn is a function of the data used and the amount of variability about the regression line. Alternatively, one may think of the $F$-test as $F=t^2$, where $t$ is equal to the difference in the two slopes divided by the standard error of the difference. Thus if we have another dataset with the same number of data points as ours but, for example, a larger (or smaller) scatter, then for this other dataset the slopes are less (or better) well defined with larger (or smaller) variance and the $F$-test statistic would generally decreases (or increases). In another situation, if we add (or remove) some data from our sample (assume the old and new samples have the same distribution), then generally the slopes will be more (or less) accurately estimated and the $F$-test statistic would increases (or decreases). We say ``generally'' because the data that is added or removed could have a large or small scatter which can, in some cases, offset the change in the amount of data. Thus the $F$-test statistic is sensitive to both the number of data points and the nature of the data.

     We have verified this by using the LMC OGLE data, as used in KN, and formulating a $F$-test for a difference in the PL slope between short ($P<10$ days) and long ($P>10$ days) period Cepheids. However we also make a cut at periods $P_1$ and $P_2$ such that $P_1< 10 < P_2$, thus decreasing the number of Cepheids available to define the slope on either side of 10 days. The results are presented in Table \ref{tabpcut} for the OGLE {\it V}-band data used in KN, as well as the MACHO {\it V}-band data as described in Section 2 (the Sample C). This table shows that the $F$-value generally decrease, or the significance of the results generally goes down (with $p$-values increase), as the number of Cepheids on either side of the break point goes down\footnote{Note that the peak of the period distribution for both samples is located around $\log(P)\sim0.5$, see Section 5.2.}. It also depends, to some extent, on the nature of the data that was removed or added. We see that when we have less than about 30 Cepheids on the long period side, the significance is reduced and the $F$-test would indicate that the data are consistent with a single line. Similarly when we have less than about 25 Cepheids on the short period side. With such few Cepheids the slopes are not well defined with large errors. This gives us confidence that when we do get a significant result, that result has some meaning.  

\subsection{Choosing $P_o$}

\begin{figure*}
       \vspace{0cm}
       \hbox{\hspace{0.2cm}\epsfxsize=8.5cm \epsfbox{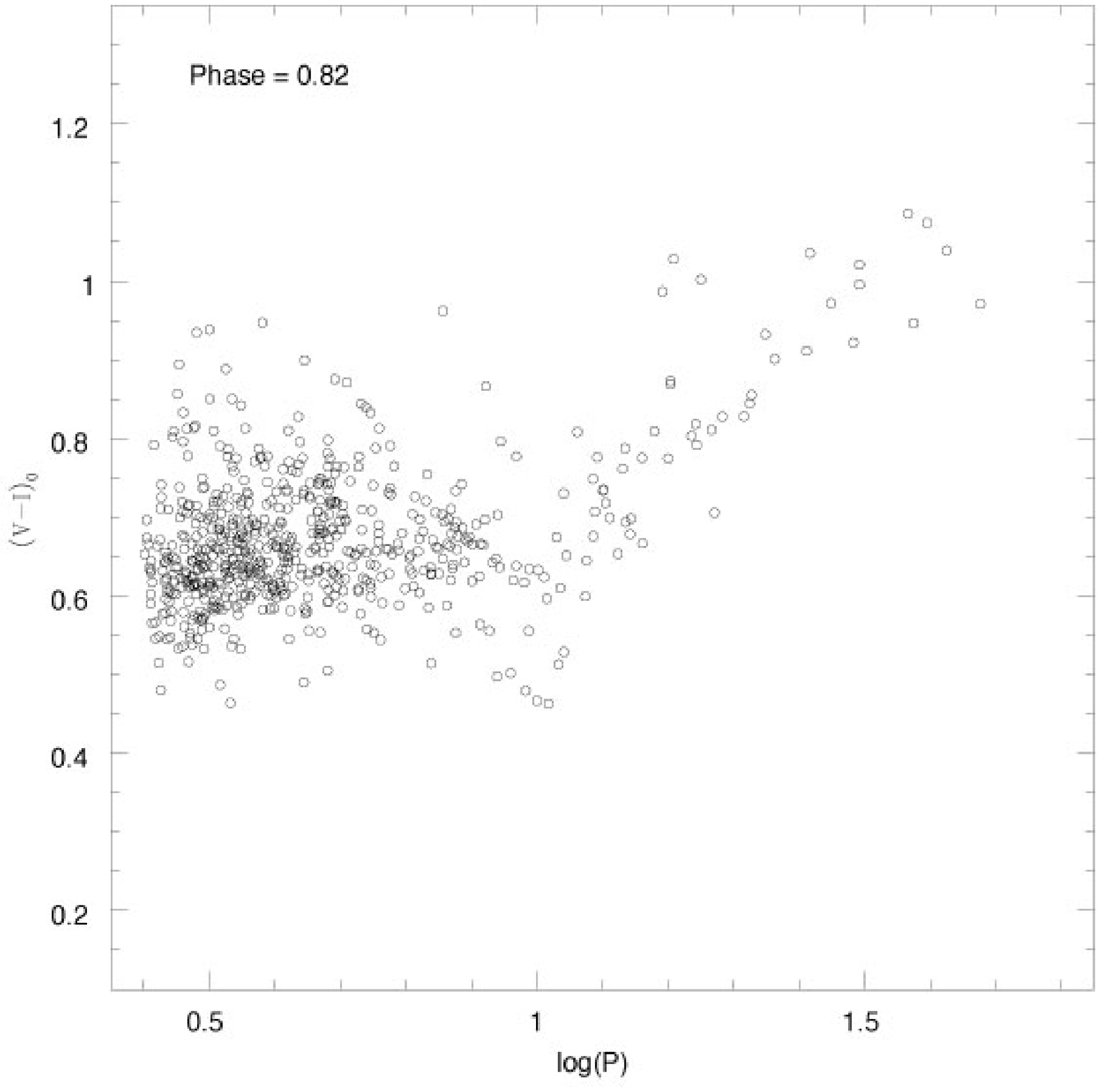}
         \epsfxsize=8.5cm \epsfbox{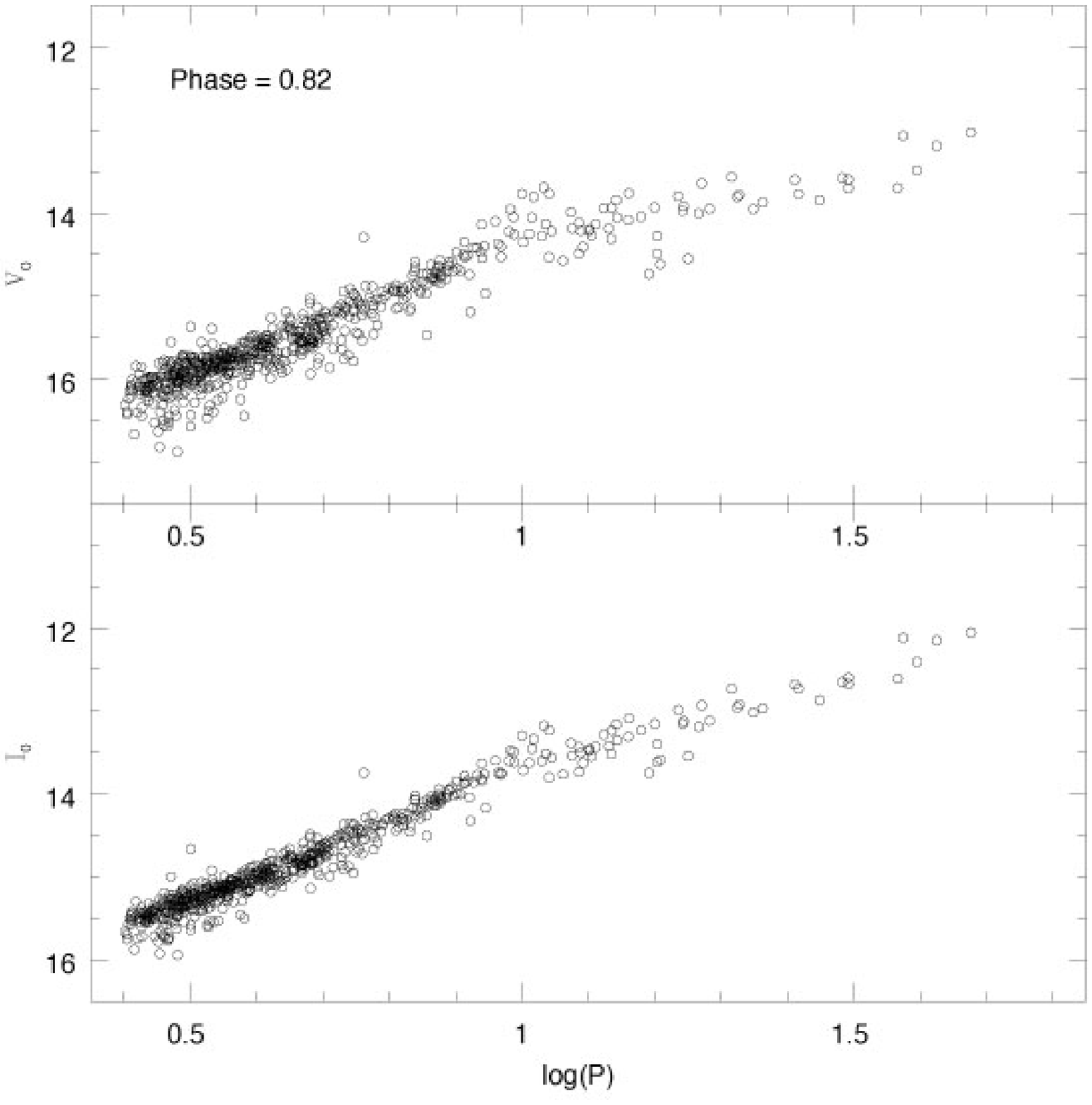}}
       \vspace{0cm}
       \caption{LMC extinction corrected PC (left panel) and PL (right panel) relations at phase of 0.82 (where phase zero = maximum light), obtained from using the OGLE photometric data. A discontinuity at/near 10days is clearly visible.}
       \label{figmulti}
     \end{figure*}

     In the recent studies of the non-linear LMC PL relations \citep{tam02,kan04,san04}, the characteristic period $P_o$ is chosen to be 10 days. \citet{san04} has discussed the significance of the discontinuity at 10 days, which will not be repeated here. However, the value of $P_o$ can be found from equation (2) by making $\log(P_o)$ another parameter in the least squares fit. To find $P_o$, we applied non-linear estimation procedures, as given in the {\tt SAS} package, to the {\it V}-band data from Sample C. The result is $\log(P_o)=0.934$, with the upper and lower 95\% confidence limits of $1.089$ and $0.778$, respectively\footnote{The result using Sample A is: $\log(P_o)=0.981$, with the 95\% confidence limits of $1.148$ and $0.815$.}. This result is fully consistent with the choice of $P_o=10$ days in \citet{tam02}, KN and \citet{san04}. Other evidence to support the choice of $P_o=10$ days include:
     
     \begin{enumerate}
     \item The amplitudes and the Fourier parameters from the light curves of the short period ($P<10$ days) Cepheids are radically different from their long period ($P>10$ days) counterparts \citep[see][and references therein]{sim81,sim85,and88,uda99,kan02,nge03,san04}, where the plots of the amplitudes and Fourier parameters as a function of $\log(P)$ show an abrupt change at 10 days. This is a well established fact in stellar pulsation. There is a well-known explanation for this change in the Fourier parameters at 10 days: the Hertzsprung progression \citep[see, e.g.][and references therein]{sim81,bon00b}. We do not imply that this theoretical explanation for the change in Fourier parameters at a period of 10 days is responsible for possible changes in the Cepheid PL relations at the same period, merely that there is a precedent in the Cepheid literature for "something happening" at a period of 10 days.
     \item In a separate paper (Ngeow \& Kanbur, 2005 -- in preparation), we study the multi-phase PC and PL relations for the OGLE LMC Cepheids. This is because the PL and PC relations at mean light \citep[as, e.g., presented in][]{tam02,kan04,san04} are the average of all other phases \citep{kan04,kan04b}. From this study, we found that both of the LMC PL and PC relations provide compelling evidence of a strong non-linearity at period of 10 days at a pulsation phase of 0.82 (where phase zero corresponds to maximum light). A preliminary result from this study is presented in Figure \ref{figmulti}, and the details will be given elsewhere.
     \item The work of \citet{tam02}, KN and \citet{san04} strongly suggests that the LMC (optical) PC relation in $(B-V)$ and/or $(V-I)$ colour is also non-linear with a break at 10 days. This can be clearly see from figure 2 of \citet{tam02}, figure 2 of KN (which also include the PC relations at maximum and minimum light), figure 1 of \citet{san04} and also the left panels of Figure \ref{figpc} in this paper.
     \item The behavior of the amplitude-colour (AC) relations for the LMC Cepheids are different for the short ($P<10$ days) and long period Cepheids at maximum, mean and minimum light (KN). This is because the PC and AC relations are connected as described in KN.
     \item The residuals plots using one-line regression shows a trend at long period range ($\log[P] \gtrsim 1.0$) for the {\it V}- and {\it R}-band PL relations (Figure \ref{figres} \& right panels of Figure \ref{fig1}), as well as the PC relations \citep[see figure 4 from][]{kan04}.   
     \item The empirical LMC instability strip, as presented in \cite{san04}, clearly show a break at a period of 10 days. The properties of the LMC instability strip, such as the ridge lines of the strip and the slope of constant-period lines, also different for the long ($\log[P]>1.0$) and short period Cepheids.
     \end{enumerate}

     \ni Hence, we follow the existing literature \citep{tam02a,kan04,san04} and adopt $P_o=10$ days in this paper. Note that due to the intrinsic dispersion of the PL relation (because of the finite width of the instability strip), the exact location of the $P_o$ is difficult to accurately determine with statistical means from the data. We emphasize that the confirmation of $P_o\simeq10$ days has to be done with the stellar pulsation modeling, which is beyond the scope of this paper. 

\section{Results}

      \begin{table}
       \centering
       \caption{The results of the one-line regressions in the form of $m=a\log(P)+b$, where $\sigma$ is the dispersion of the regression.}
       \label{tab1line}
       \begin{tabular}{lccc} \hline 
         Bandpass  & $a$ & $b$ & $\sigma$ \\     
         \hline 
         \multicolumn{4}{c}{Sample A, $N=1330$} \\
         $V\dots$ & $-2.669\pm0.031$ & $17.074\pm0.020$ & 0.285 \\
         $R\dots$ & $-2.826\pm0.027$ & $16.860\pm0.018$ & 0.244 \\
         $J\dots$ & $-3.080\pm0.022$ & $16.277\pm0.014$ & 0.130 \\
         $H\dots$ & $-3.175\pm0.020$ & $16.046\pm0.013$ & 0.114 \\
         $K\dots$ & $-3.156\pm0.020$ & $15.973\pm0.014$ & 0.129 \\
         \multicolumn{4}{c}{Sample B, $N=1322$} \\
         $V\dots$ & $-2.672\pm0.031$ & $17.066\pm0.020$ & 0.257 \\
         $R\dots$ & $-2.828\pm0.027$ & $16.853\pm0.018$ & 0.220 \\
         $J\dots$ & $-3.080\pm0.022$ & $16.275\pm0.014$ & 0.124 \\
         $H\dots$ & $-3.174\pm0.020$ & $16.044\pm0.013$ & 0.112 \\
         $K\dots$ & $-3.156\pm0.020$ & $15.972\pm0.014$ & 0.128 \\
         \multicolumn{4}{c}{Sample C, $N=1216$} \\
         $V\dots$ & $-2.699\pm0.035$ & $17.087\pm0.023$ & 0.259 \\
         $R\dots$ & $-2.852\pm0.030$ & $16.871\pm0.020$ & 0.222 \\
         $J\dots$ & $-3.088\pm0.024$ & $16.281\pm0.016$ & 0.124 \\
         $H\dots$ & $-3.169\pm0.022$ & $16.040\pm0.015$ & 0.110 \\
         $K\dots$ & $-3.159\pm0.021$ & $15.974\pm0.015$ & 0.124 \\
         \hline
       \end{tabular}
     \end{table}

     In Table \ref{tab1line} - \ref{tabquad}, we present the fitted results for one-line, two-lines, piecewise continuous and quadratic regressions, respectively, for each of the three samples considered in this paper. It is worthwhile to point out that even though the dispersion values
     (given by  $\sqrt{\sum [m(i) - \hat{m}(i)]^2/(N-n)}$, which is not the same as the $SSE$ that enters the chi-square test and the $F$-test) given in the last column of these tables are comparable to each other\footnote{We thank the anonymous referee to point this out.}, though with a slight improvement in Table \ref{tab2line} \& \ref{tabpie} as compared to Table \ref{tab1line}. This, however, does not mean that the four regression models describe the data equally well. The $SSE$ is approximately the dispersion multiplied by the number of data points so differences between the models are more apparent. Further, a statistical test based purely on a comparison of the dispersion does not have the required power and is not a good statistical test to apply to our data. This is due to the following reasons: (a) the dispersion is dominated by the intrinsic dispersion ($\sigma_{IS}$) because of the existence of the instability strip; (b) the majority of the Cepheids are in the short period range which dominates the dispersion; and (c) more importantly, the non-linear signature of the mean light LMC PL relation is not obvious as shown in, e.g., Figure \ref{figmulti}, or with a large quadratic term ($\gamma$) in equation (3). Therefore, it is necessary to apply the chi-square test and the $F$-test to detect such non-linear signature from our data.

     The extinction corrected {\it VRJHK}-band PL relations are presented in the left panels of Figure \ref{fig1} (from Sample C for illustrative purposes, together with the fitted two-lines regressions. The PL relations from the other two samples, as well as other forms of regressions, are similar to Figure \ref{fig1}). From Figure \ref{fig1}, the {\it V}- and {\it R}-band PL relations imply that there is a break or sharp non-linearity at/around 10 days, marginally in the {\it J}-band PL relation, but not obvious in the {\it H}- and {\it K}-band PL relations. The right panels of Figure \ref{fig1} compare the residuals of the fit as a function of period using one-line and two-line regressions (again the residual plots from the two other forms of regressions are very similar to these plots). From the residual plots, both of the {\it V}- and {\it R}-band residuals show a trend at $\log(P)>1.0$ when using the one-line regression. These trends are removed if two-line regressions are used to fit the data. These findings also support a break in the PL relation at a period of 10 days. Similarly, the {\it J}-band residuals show a marginal trend, which is not obvious in the {\it H}- and {\it K}-band. 

     \begin{figure*}
       \vspace{0cm}
       \hbox{\hspace{2.0cm}\epsfxsize=7.0cm \epsfbox{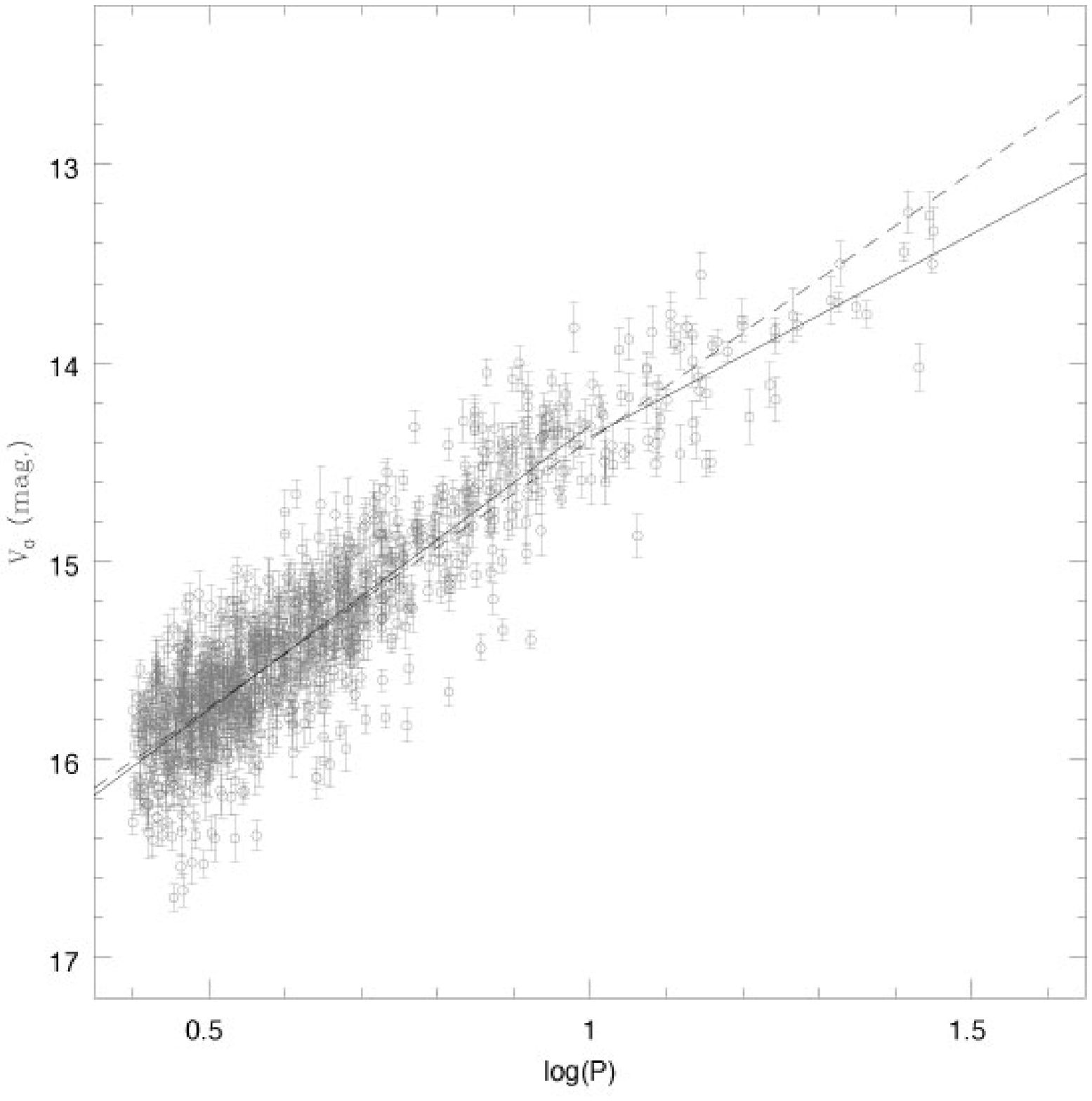}
         \epsfxsize=7.0cm \epsfbox{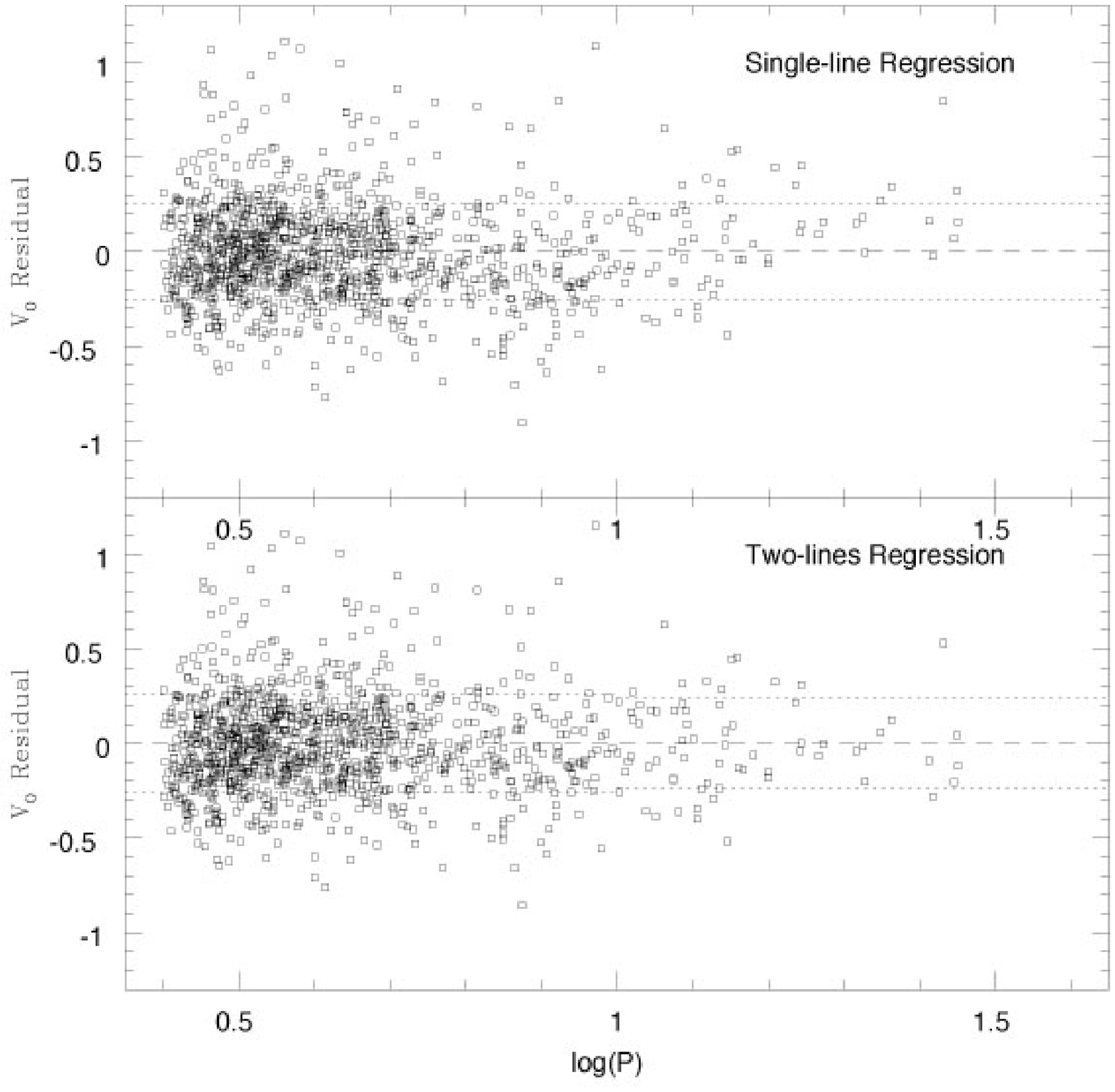}}
       \hbox{\hspace{2.0cm}\epsfxsize=7.0cm \epsfbox{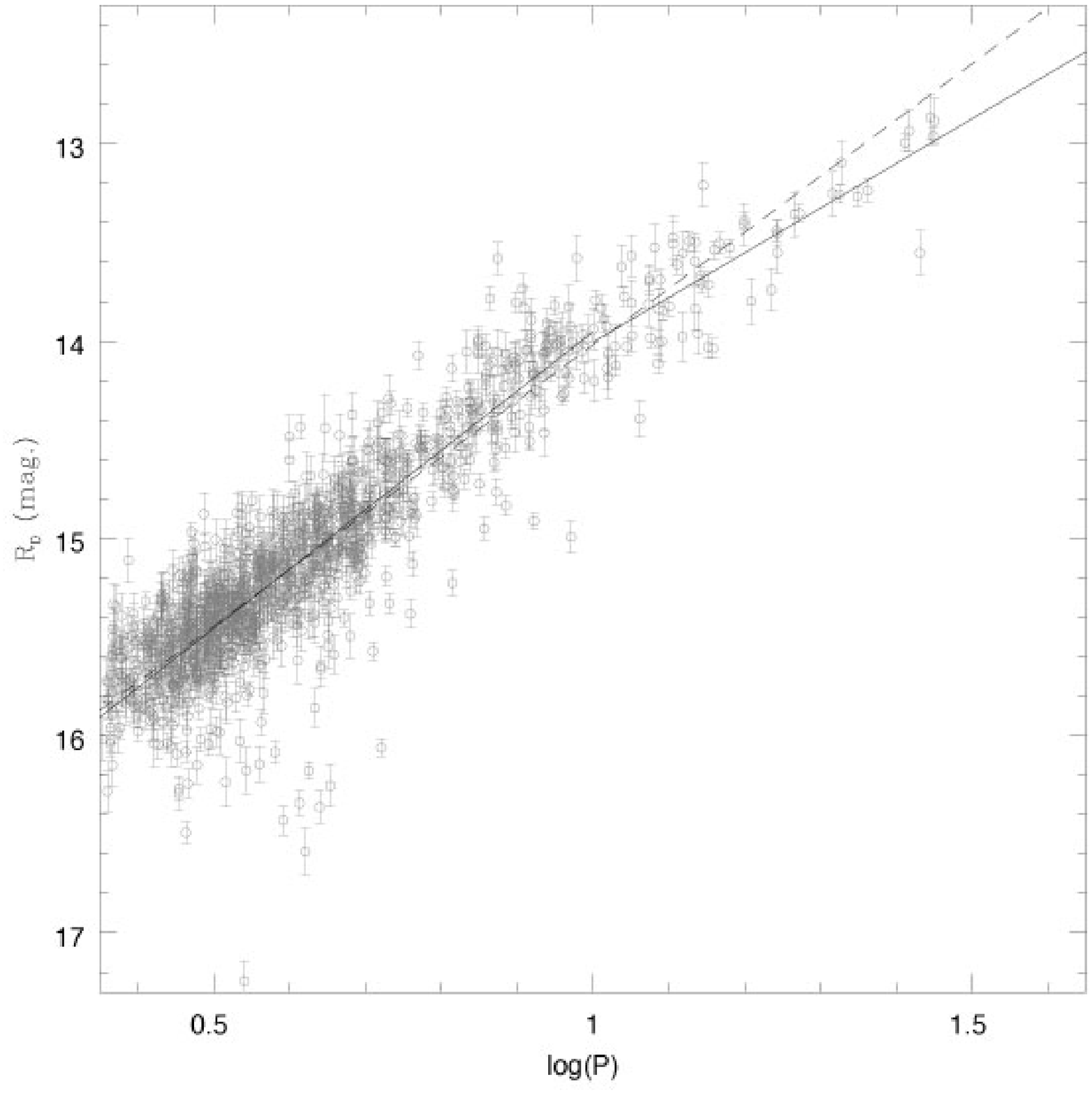}
         \epsfxsize=7.0cm \epsfbox{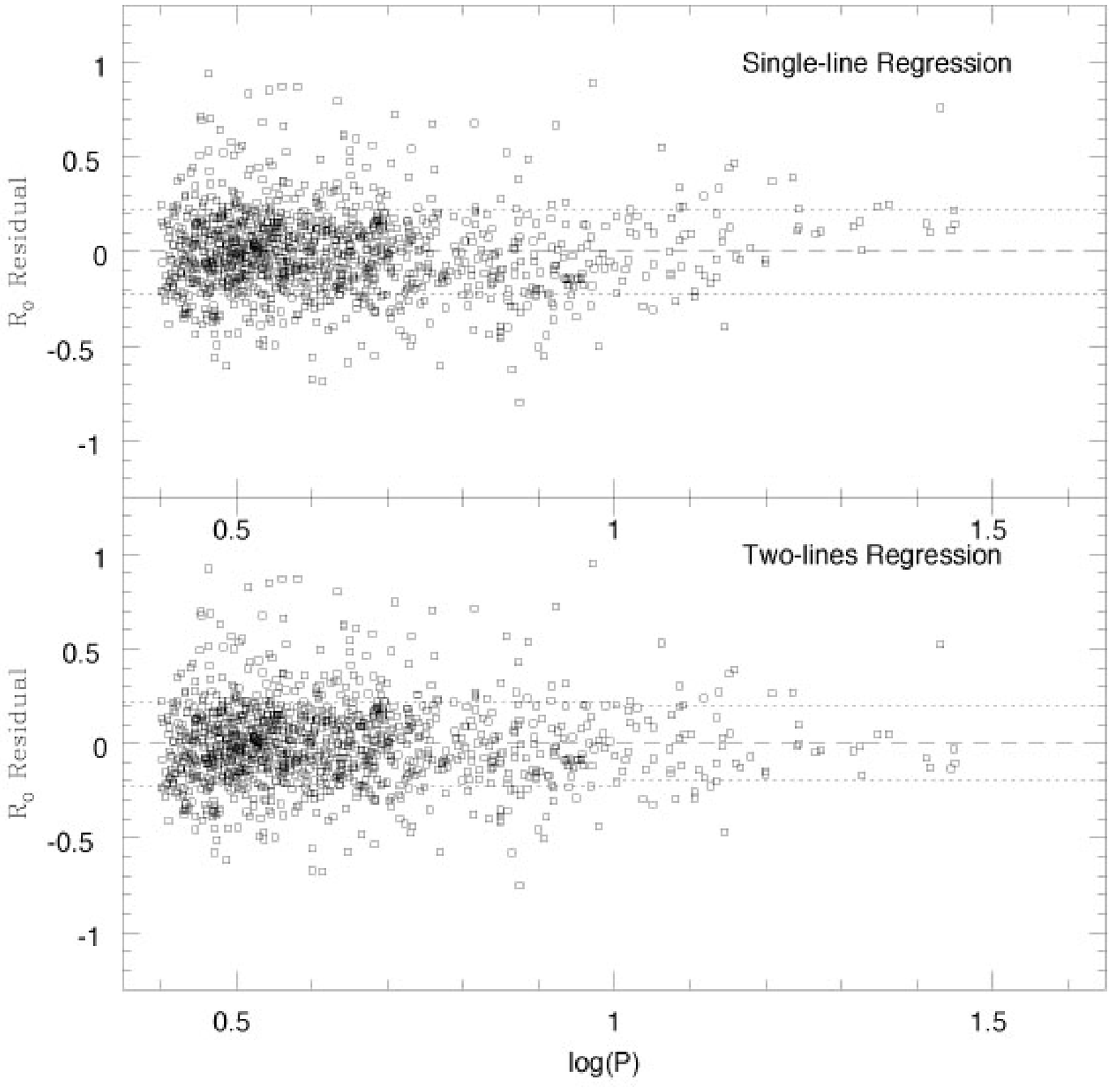}}
       \hbox{\hspace{2.0cm}\epsfxsize=7.0cm \epsfbox{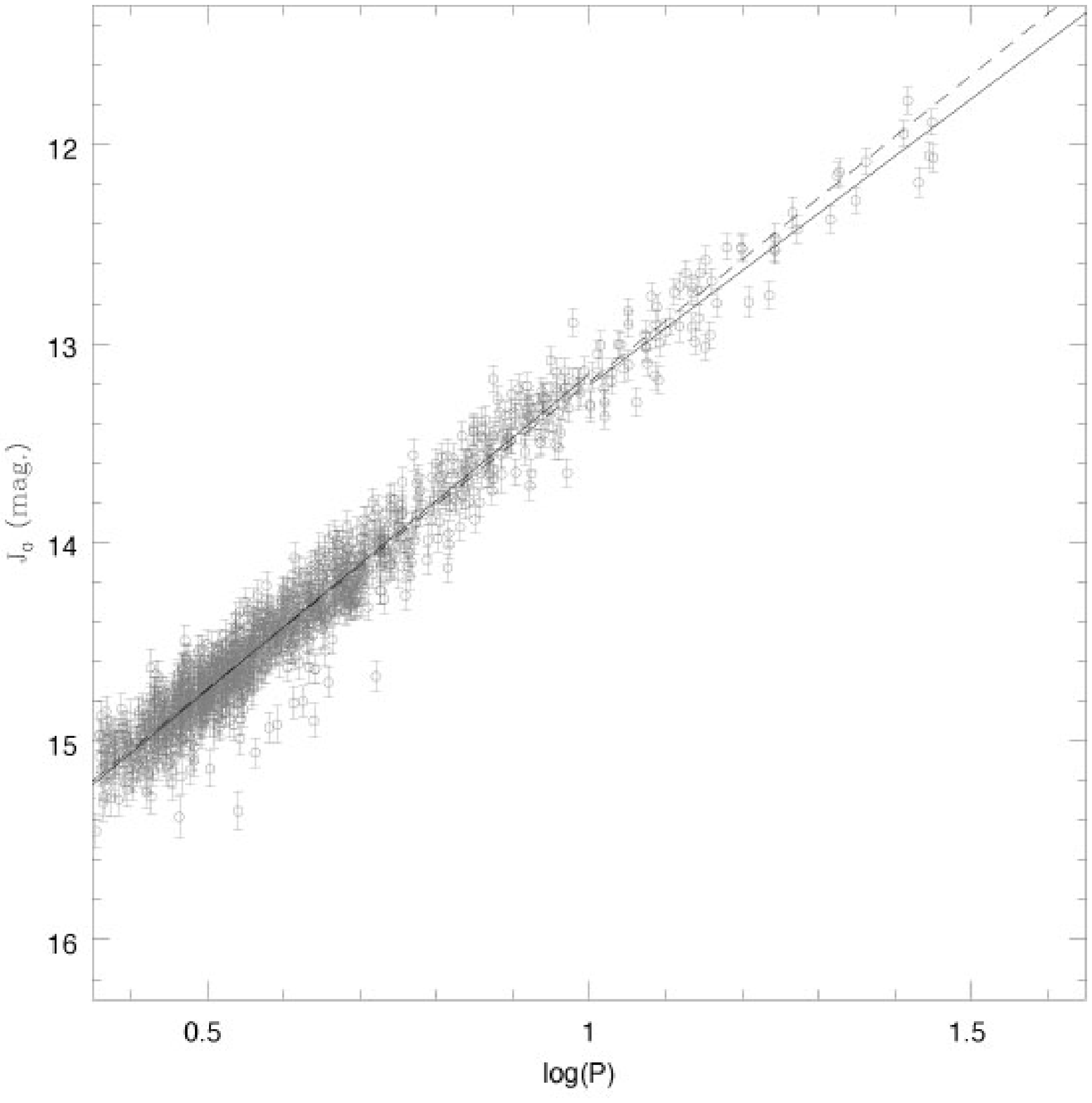}
         \epsfxsize=7.0cm \epsfbox{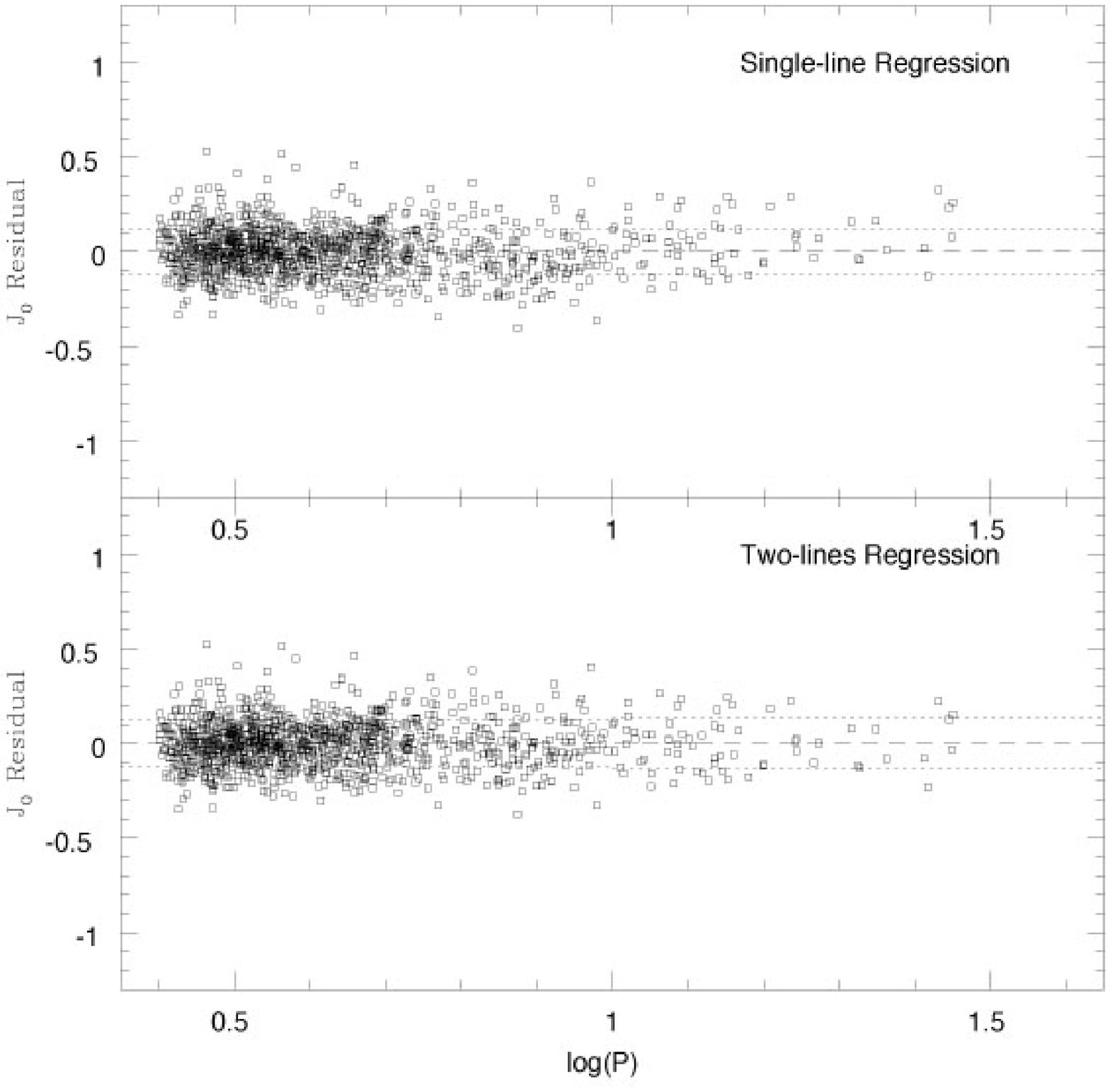}}
       \vspace{0cm}
       \caption{{\it Left (a)}: The extinction corrected {\it VRJHK}-band PL relations for Sample C. The dashed and solid lines are the fitted one-line and two-line regressions, respectively. The error bars are the quadrature sum of the photometric errors in mean magnitudes ($\sigma_{\lambda}$), the errors from the random phase corrections that are only applicable to {\it JHK}-bands ($\sigma_{rp}$) and the errors from the extinction ($\sigma_{A_{\lambda}}$). {\it Right (b)}: The residuals from using one-line (top panel) and two-line (bottom panel) regression fits. The dashed line indicates the zero residual. The dotted lines are the $1\sigma$ dispersions.}
       \label{fig1}
     \end{figure*}

     \begin{figure*}
       \vspace{0cm}
       \hbox{\hspace{2.0cm}\epsfxsize=7.0cm \epsfbox{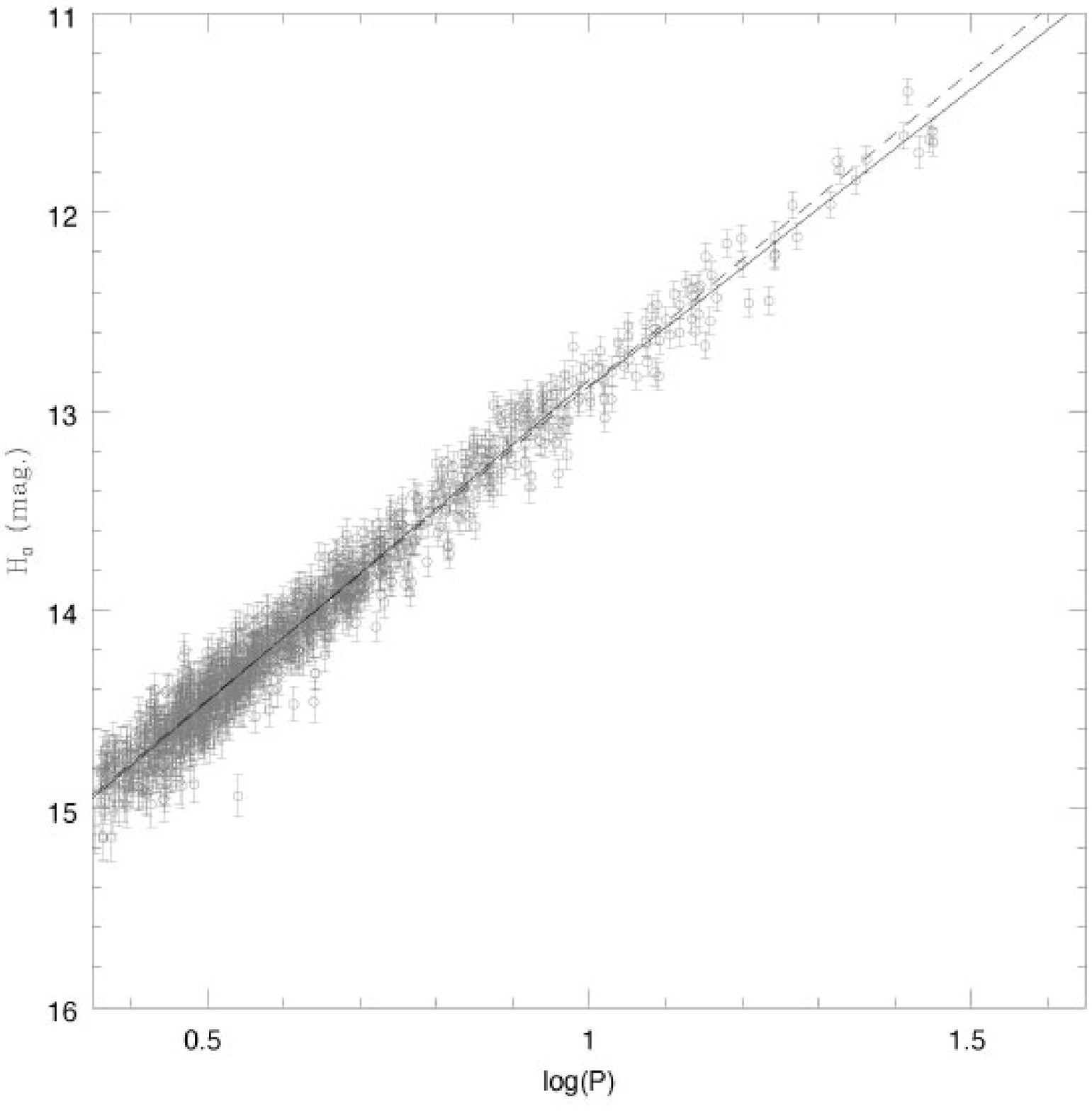}
         \epsfxsize=7.0cm \epsfbox{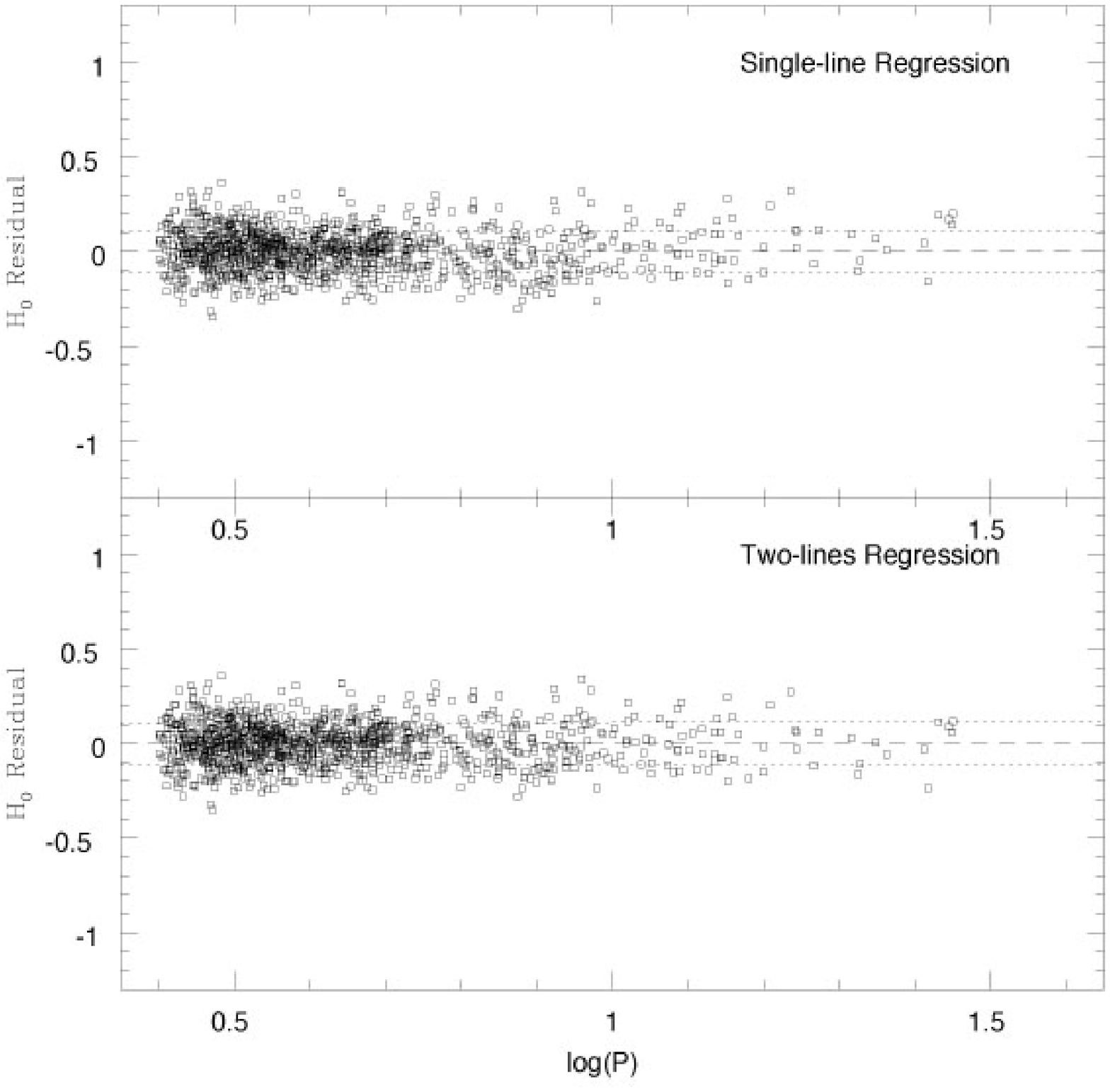}}
       \hbox{\hspace{2.0cm}\epsfxsize=7.0cm \epsfbox{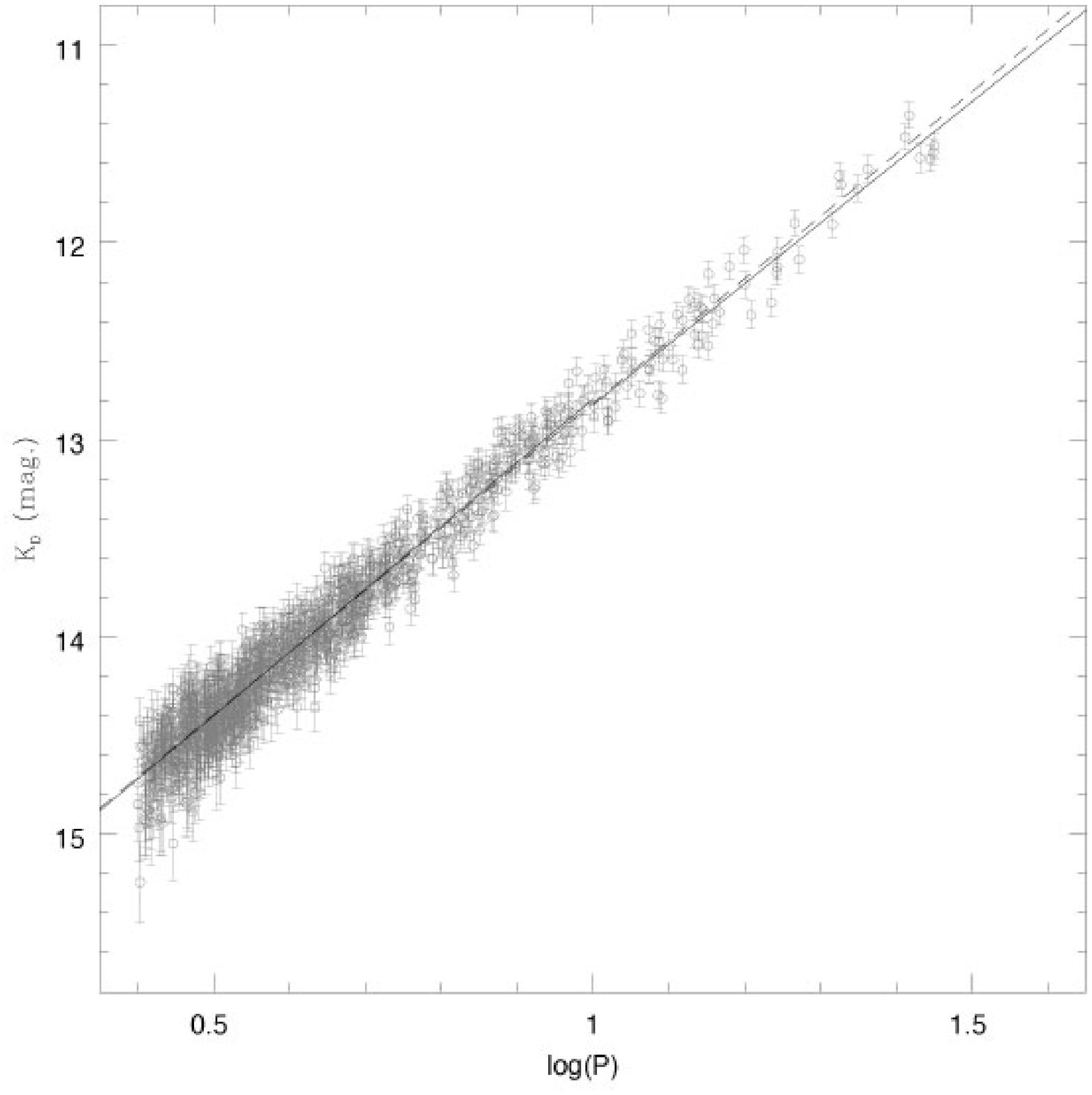}
         \epsfxsize=7.0cm \epsfbox{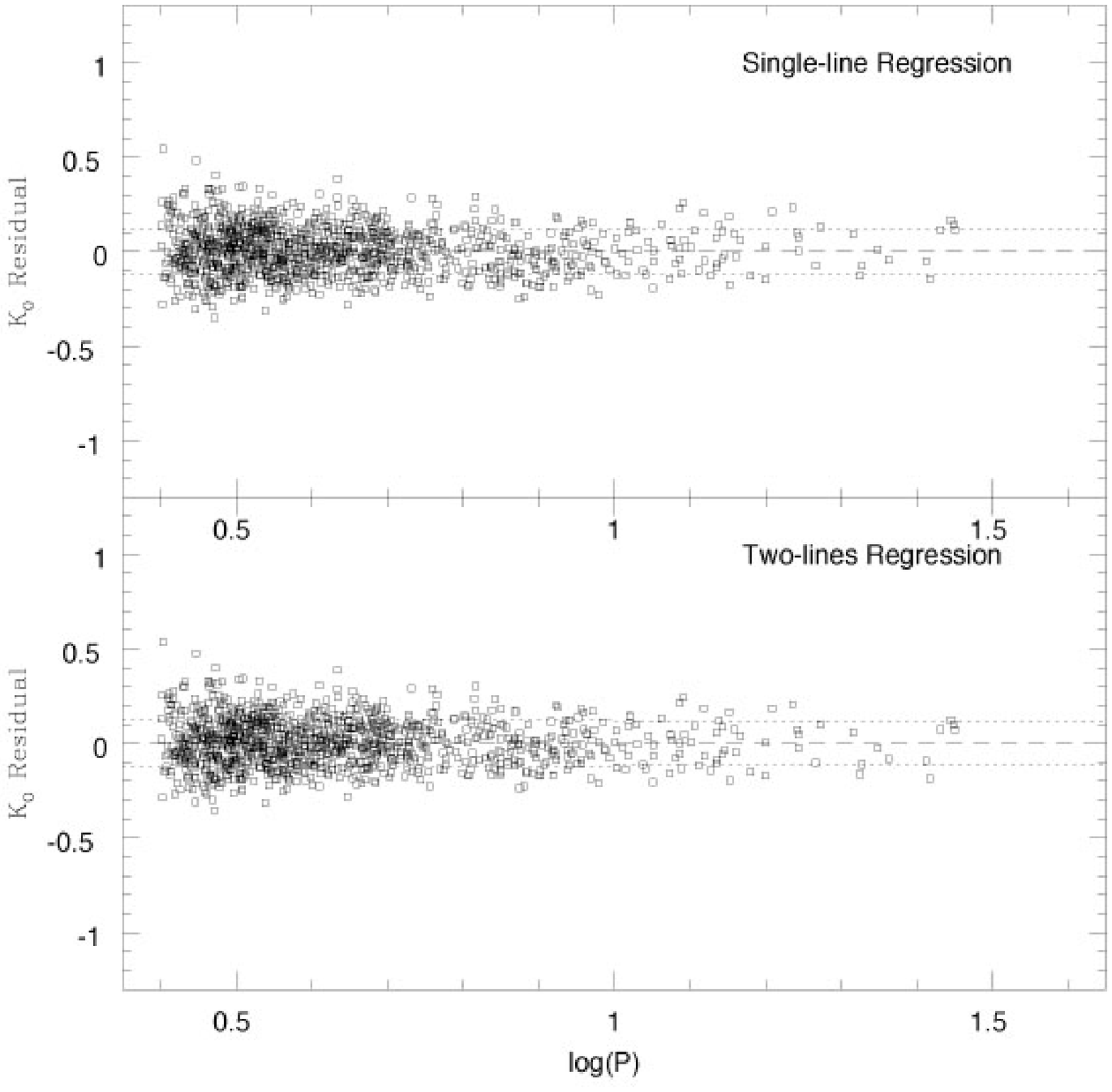}}
       \vspace{0cm}
       \addtocounter{figure}{-1}
       \caption{\emph{continued}.}
     \end{figure*}

     Both of the chi-square test and the $F$-test were then applied to the samples. The results from the chi-square test and the $F$-test are summarized in Table \ref{tabch2a} \& \ref{tabfa}, respectively. These results can be compared with Table \ref{tabp} to decide if the null hypothesis can be rejected. If the $\chi^2$ and/or the $F$ values given in Table \ref{tabch2a} \& \ref{tabfa} are smaller than the values corresponding to $p=0.05$, then the underlying PL relation is linear. In contrast, if the $\chi^2$ and/or the $F$ values in these tables are larger than the values corresponding to $p=0.01$, then the underlying PL relation is non-linear. For the the $\chi^2$ and/or the $F$ values that fall in-between the values corresponding to $p=0.05$ and $p=0.01$, we classify the PL relation to be marginally linear. In short, from Table \ref{tabch2a} \& \ref{tabfa} we found that:

     \begin{table*}
       \centering
       \caption{The results of the two-line regressions in the form of $m=a_{(S,L)}\log(P)+b_{(S,L)}$, where $\sigma_{(S,L)}$ is the dispersion of the regression. The subscripts $_{S,L}$ are refereed to the short ($P<10$ days) and long period Cepheids, respectively.}
       \label{tab2line}
       \begin{tabular}{lcccccc} \hline 
         Bandpass  & $a_S$ & $b_S$ & $\sigma_S$ & $a_L$ & $b_L$ & $\sigma_L$\\     
         \hline 
         \multicolumn{7}{c}{Sample A, $N_S=1261$, $N_L=69$} \\
         $V\dots$ & $-2.771\pm0.041$ & $17.130\pm0.025$ & 0.286 & $-2.039\pm0.222$ & $16.409\pm0.258$ & 0.238 \\
         $R\dots$ & $-2.919\pm0.036$ & $16.911\pm0.022$ & 0.245 & $-2.263\pm0.193$ & $16.267\pm0.224$ & 0.200 \\
         $J\dots$ & $-3.143\pm0.029$ & $16.312\pm0.018$ & 0.129 & $-2.884\pm0.151$ & $16.093\pm0.175$ & 0.132 \\
         $H\dots$ & $-3.234\pm0.027$ & $16.078\pm0.017$ & 0.113 & $-3.003\pm0.134$ & $15.883\pm0.155$ & 0.117 \\
         $K\dots$ & $-3.193\pm0.028$ & $15.994\pm0.017$ & 0.129 & $-3.080\pm0.125$ & $15.904\pm0.145$ & 0.113 \\
         \multicolumn{7}{c}{Sample B, $N_S=1253$, $N_L=69$} \\
         $V\dots$ & $-2.787\pm0.041$ & $17.129\pm0.025$ & 0.256 & $-2.039\pm0.222$ & $16.409\pm0.258$ & 0.238 \\
         $R\dots$ & $-2.933\pm0.036$ & $16.910\pm0.022$ & 0.220 & $-2.263\pm0.193$ & $16.267\pm0.224$ & 0.200 \\
         $J\dots$ & $-3.148\pm0.029$ & $16.312\pm0.018$ & 0.123 & $-2.884\pm0.151$ & $16.093\pm0.175$ & 0.132 \\
         $H\dots$ & $-3.235\pm0.027$ & $16.078\pm0.017$ & 0.110 & $-3.003\pm0.134$ & $15.883\pm0.155$ & 0.117 \\
         $K\dots$ & $-3.195\pm0.028$ & $15.994\pm0.017$ & 0.129 & $-3.080\pm0.125$ & $15.904\pm0.145$ & 0.113 \\
         \multicolumn{7}{c}{Sample C, $N_S=1147$, $N_L=69$} \\
         $V\dots$ & $-2.867\pm0.048$ & $17.184\pm0.030$ & 0.257 & $-2.039\pm0.222$ & $16.409\pm0.258$ & 0.238 \\
         $R\dots$ & $-3.005\pm0.042$ & $16.959\pm0.026$ & 0.221 & $-2.263\pm0.193$ & $16.267\pm0.224$ & 0.200 \\
         $J\dots$ & $-3.179\pm0.033$ & $16.333\pm0.021$ & 0.122 & $-2.884\pm0.151$ & $16.093\pm0.175$ & 0.132 \\
         $H\dots$ & $-3.239\pm0.030$ & $16.081\pm0.019$ & 0.109 & $-3.003\pm0.134$ & $15.883\pm0.155$ & 0.117 \\
         $K\dots$ & $-3.207\pm0.030$ & $16.002\pm0.019$ & 0.125 & $-3.080\pm0.125$ & $15.904\pm0.145$ & 0.113 \\
         \hline
       \end{tabular}
     \end{table*}

     \begin{table*}
       \centering
       \caption{The results of the piecewise continuous regressions in the form of $m=c + d\times \mathrm{min}[\log(P),1.0)]+e\times \mathrm{max}[0,\log(P)-1.0]$, where $\sigma$ is the dispersion of the regression. }
       \label{tabpie}
       \begin{tabular}{lcccc} \hline 
         Bandpass  & $c$ & $d$ & $e$ & $\sigma$ \\     
         \hline 
         \multicolumn{5}{c}{Sample A, $N=1330$} \\
         $V\dots$ & $17.128\pm0.024$ & $-2.767\pm0.038$ & $-2.000\pm0.155$ & 0.284 \\
         $R\dots$ & $16.908\pm0.021$ & $-2.915\pm0.033$ & $-2.219\pm0.135$ & 0.243 \\
         $J\dots$ & $16.305\pm0.017$ & $-3.130\pm0.027$ & $-2.751\pm0.106$ & 0.129 \\
         $H\dots$ & $16.072\pm0.016$ & $-3.221\pm0.025$ & $-2.889\pm0.094$ & 0.113 \\
         $K\dots$ & $15.989\pm0.016$ & $-3.184\pm0.025$ & $-3.004\pm0.089$ & 0.129 \\
         \multicolumn{5}{c}{Sample B, $N=1322$} \\
         $V\dots$ & $17.125\pm0.024$ & $-2.778\pm0.038$ & $-1.944\pm0.155$ & 0.255 \\
         $R\dots$ & $16.906\pm0.021$ & $-2.925\pm0.033$ & $-2.170\pm0.135$ & 0.219 \\
         $J\dots$ & $16.304\pm0.017$ & $-3.133\pm0.027$ & $-2.735\pm0.106$ & 0.124 \\
         $H\dots$ & $16.071\pm0.016$ & $-3.222\pm0.025$ & $-2.880\pm0.094$ & 0.111 \\
         $K\dots$ & $15.989\pm0.016$ & $-3.185\pm0.025$ & $-2.999\pm0.089$ & 0.128 \\
         \multicolumn{5}{c}{Sample C, $N=1216$} \\
         $V\dots$ & $17.172\pm0.028$ & $-2.846\pm0.044$ & $-1.863\pm0.157$ & 0.256 \\
         $R\dots$ & $16.948\pm0.024$ & $-2.985\pm0.038$ & $-2.098\pm0.137$ & 0.219 \\
         $J\dots$ & $16.321\pm0.019$ & $-3.157\pm0.030$ & $-2.708\pm0.107$ & 0.123 \\
         $H\dots$ & $16.072\pm0.018$ & $-3.223\pm0.028$ & $-2.879\pm0.095$ & 0.109 \\
         $K\dots$ & $15.995\pm0.018$ & $-3.194\pm0.028$ & $-2.988\pm0.090$ & 0.124 \\
         \hline
       \end{tabular}
     \end{table*}

     \begin{table*}
       \centering
       \caption{The results of the quadratic regressions in the form of $m=\alpha + \beta\log(P) + \gamma [\log(P)]^2$, where $\sigma$ is the dispersion of the regression. }
       \label{tabquad}
       \begin{tabular}{lcccc} \hline 
         Bandpass  & $\alpha$ & $\beta$ & $\gamma$ & $\sigma$ \\     
         \hline 
         \multicolumn{5}{c}{Sample A, $N=1330$} \\
         $V\dots$ & $17.126\pm0.050$ & $-2.829\pm0.144$ & $0.110\pm0.097$ & 0.285 \\
         $R\dots$ & $16.914\pm0.044$ & $-2.993\pm0.125$ & $0.115\pm0.084$ & 0.244 \\
         $J\dots$ & $16.336\pm0.036$ & $-3.259\pm0.102$ & $0.123\pm0.068$ & 0.130 \\
         $H\dots$ & $16.125\pm0.035$ & $-3.412\pm0.098$ & $0.159\pm0.065$ & 0.114 \\
         $K\dots$ & $15.996\pm0.037$ & $-3.222\pm0.102$ & $0.043\pm0.066$ & 0.129 \\
         \multicolumn{5}{c}{Sample B, $N=1322$} \\
         $V\dots$ & $17.160\pm0.050$ & $-2.961\pm0.144$ & $0.200\pm0.097$ & 0.257 \\
         $R\dots$ & $16.944\pm0.044$ & $-3.108\pm0.125$ & $0.193\pm0.084$ & 0.220 \\
         $J\dots$ & $16.346\pm0.036$ & $-3.296\pm0.103$ & $0.148\pm0.068$ & 0.124 \\
         $H\dots$ & $16.131\pm0.035$ & $-3.434\pm0.098$ & $0.174\pm0.065$ & 0.111 \\
         $K\dots$ & $16.000\pm0.037$ & $-3.236\pm0.102$ & $0.052\pm0.066$ & 0.128 \\
         \multicolumn{5}{c}{Sample C, $N=1216$} \\
         $V\dots$ & $17.345\pm0.072$ & $-3.435\pm0.197$ & $0.477\pm0.126$ & 0.258 \\
         $R\dots$ & $17.114\pm0.063$ & $-3.545\pm0.171$ & $0.449\pm0.109$ & 0.221 \\
         $J\dots$ & $16.427\pm0.050$ & $-3.502\pm0.136$ & $0.267\pm0.086$ & 0.123 \\
         $H\dots$ & $16.159\pm0.046$ & $-3.505\pm0.124$ & $0.215\pm0.078$ & 0.109 \\
         $K\dots$ & $16.031\pm0.046$ & $-3.317\pm0.123$ & $0.100\pm0.077$ & 0.124 \\
         \hline
       \end{tabular}
     \end{table*}

     \begin{enumerate}
     \item For $H_A=$ two-line and piecewise continuous regressions, both of the chi-square test and the $F$-test show that the {\it VRJ} \& {\it H}-band data (in all three samples) are more consistent with the two-line regression, and the null hypothesis with one-line regression can be rejected with very high confidence level ($>99.9$\%).  In contrast, the {\it K}-band data (in all three samples) is consistent with the one-line regression from the chi-square test. The $F$-test results for the {\it K}-band data also show that the {\it K}-band PL relation is marginally linear\footnote{Note that for the {\it K}-band results, the $F$-values for $H_A=$piecewise continuous regression are larger than the $F$-values from $H_A=$two-line regression. Hence the {\it K}-band PL relation will be more consistent with the piecewise continuous regression if $p=0.01$ is chosen and vice versa.}. 
     \item For $H_A=$ quadratic regression, the results are quite different to the two other types of regression. The linear/non-linear behavior is quite different in {\it VRJ}- and {\it H}-band data across the three samples. From the chi-square test, the {\it VRJ}-band PL relations are linear in Sample A, then become marginally linear in Sample B and non-linear in Sample C. The {\it H}-band PL relation is marginally linear in Sample A but non-linear in both of the Sample B \& C from the chi-square test. In terms of the $F$-test, Sample A shows that the {\it V}- \& {\it R}-band PL relations are linear, marginally linear for the {\it J}-band PL relation and non-linear for the {\it H}-band PL relation. In Sample B, the {\it V}-band PL relation is still linear, but the {\it R}-band PL relation becomes marginally linear, and both of the {\it J}- and {\it H}-band PL relations are non-linear. The {\it VRJ}- and {\it H}-band PL relations become non-linear in Sample C. In contrast, the {\it K}-band PL relations are linear in all three samples from both of the chi-square test and the $F$-test.        
     \end{enumerate}

     In terms of the sample selection, the results from Sample A are not a good choice as the final adopted results. This is because Sample A suffers from obvious outliers as shown in Figure \ref{figoutlier}. Further, these outliers have made the quadratic fit for the {\it VRJ}-band PL relations to be more linear (as the quadratic term in equation [3], $\gamma$, is consistent with zero, see Table \ref{tabquad}), in contrast to the results from the two-line or piecewise continuous regressions. Hence the results from Sample A is not considered further. Both of the Sample B and Sample C are freed from the obvious outliers, with the difference that Sample B includes Cepheids with $\log(P)<0.4$ but Sample C excludes these very short period Cepheids. The statistical tests with two-line and piecewise continuous regressions show consistent results for both samples, but the results with quadratic regression are slightly different as mentioned previously. This implies that neglecting Cepheids with $\log(P)<0.4$ does not influence our results. We favor the results from Sample C because there are some physical reasons to exclude the Cepheids with $\log(P)<0.4$, as discussed in Section 2 (the contamination from overtone Cepheids and the possible changes of slope due to evolutionary effect). 

     Among the three full non-linear models, the true form of the non-linearity is difficult to distinguished statistically from the data. This has to be done together with the theoretical calculations, which is beyond the scope of this paper. However, we favor both two-line and piecewise continuous regression model over the polynomial fit because of the following reasons: (a) the residual plots in Figure \ref{fig1} suggest that the two-line/piecewise continuous regression is more applicable; (b) the (optical) PC relations are more consistent with two-line or piecewise continuous regression (see left panels of Figure \ref{figpc} below and the PC relation presented in \citealt{tam02a,kan04,san04}); (c) a simple period-mean density relation predicts a PL relation linear in $\log(P)$ but not in $[\log(P)]^2$; and (d) Figure \ref{figmulti} shows that the two-line or the piecewise continuous regressions are more appropriate to describe the data. Nevertheless it is clear this question is rather moot since a two-line or a piecewise continuous model with two separate slopes for short and long period Cepheids approximates a quadratic PL relation. For the (4-parameter) two-line regression model vs. the piecewise continuous regression model, it is true that either model may be appropriate. Again this has to wait for the future theoretical calculations to answer this problem.

     In short, from the above discussion, the main results we found from this study are:

     \emph{Based on the data used in this study, the LMC {\it V}- \& {\it R}-band relations are non-linear around a period of 10 days. Assuming that veracity of the random phase corrections used for the {\it JHK} data, we also find the {\it J}- \& {\it H}-band PL relations are also non-linear across this same period but the {\it K}-band PL relation is (marginally) linear.}

\section{Conclusion and Discussion}

     By applying the statistical tests to the {\it VRJHK}-band PL relations obtained from the MACHO and 2MASS data, we find that the {\it VRJH}-band PL relations show strong evidence that they are non-linear while the {\it K}-band PL relation is (marginally) consistent with a linear PL relation. The non-linearity of the extinction corrected {\it V}-band PL relations as seen from two independent datasets, one from the OGLE data \citep{tam02a,kan04,san04} and another from this study, strongly indicates that this non-linearity is real and not due to the artifacts of, for example, the extinction error and/or the photometric reductions. Note that \citet{per04} suggested the NIR PL relations (based on $88$ LMC Cepheids) are linear. Their results may well be influenced by the small number of short period Cepheids ($18$) in their sample. Recall that Table \ref{tabpcut} indicates how small numbers of short and/or long period Cepheids can lead to an $F$-test indicating consistency with one-line regression. We can apply the $F$-test to the 88 Cepheids (after the extinction correction) given in \citet{per04}, including the 18 short period Cepheids. The $F$-test results are: $F_J=2.825$ ($p=0.065$), $F_H=2.356$ ($p=0.101$) and $F_K=2.215$ ($p=0.116$). These results indicate that the corresponding PL relations are linear, as expected (from Table \ref{tabpcut}), and as reported in \citet{per04}, though the result for the {\it J}-band is marginal. The reason for these non-significant results is because of the small number of short period Cepheids (however, see Section 5.2), leading to a short period slope that has a large error. However, we also see that our $F$-test results for this dataset has decreasing significance as the wavelength increases. This is exactly what we see in our results for the MACHO/2MASS {\it JHK} data.

\begin{figure*}
       \vspace{0cm}
       \hbox{\hspace{0.2cm}\epsfxsize=8.5cm \epsfbox{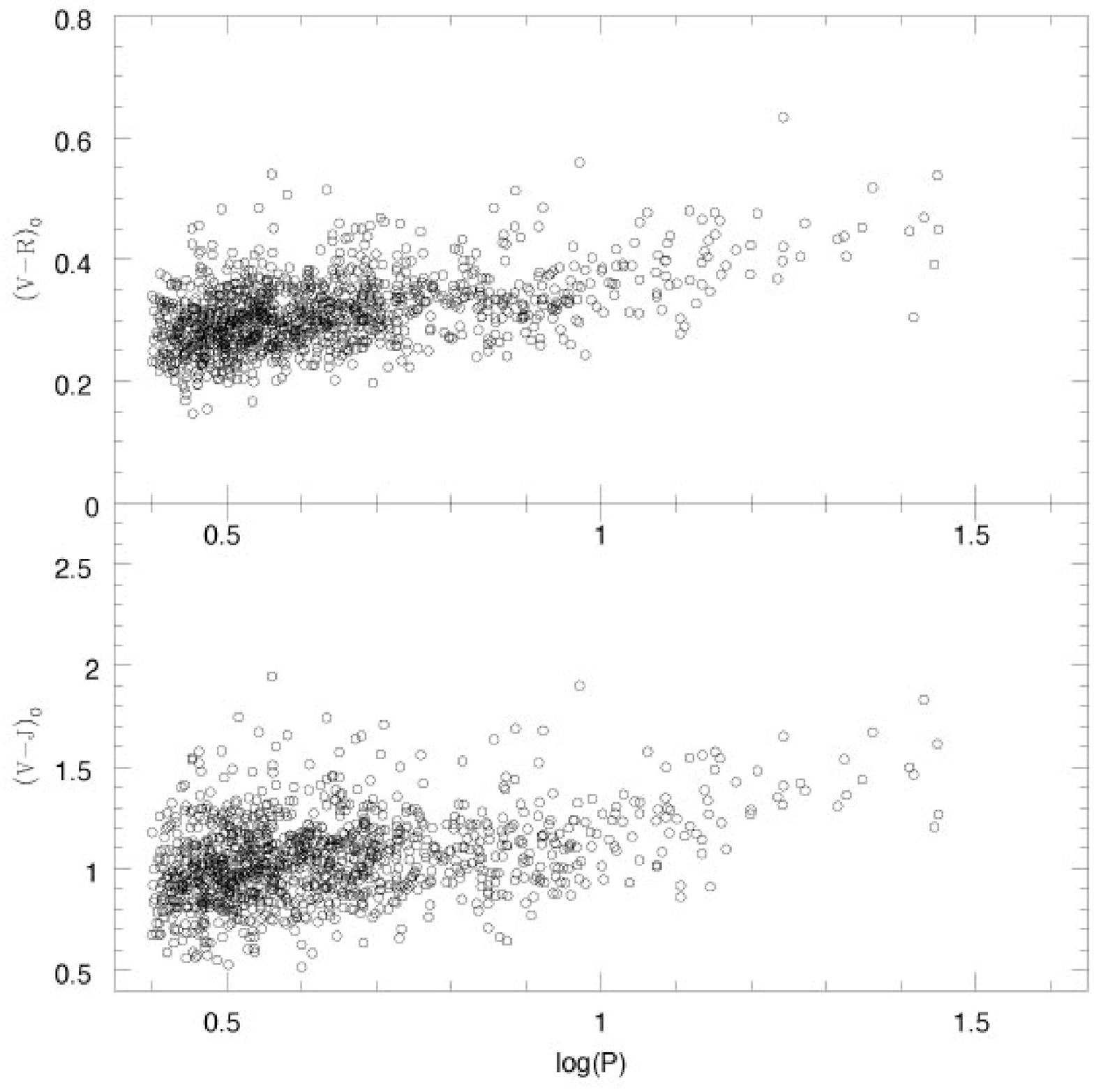}
         \epsfxsize=8.5cm \epsfbox{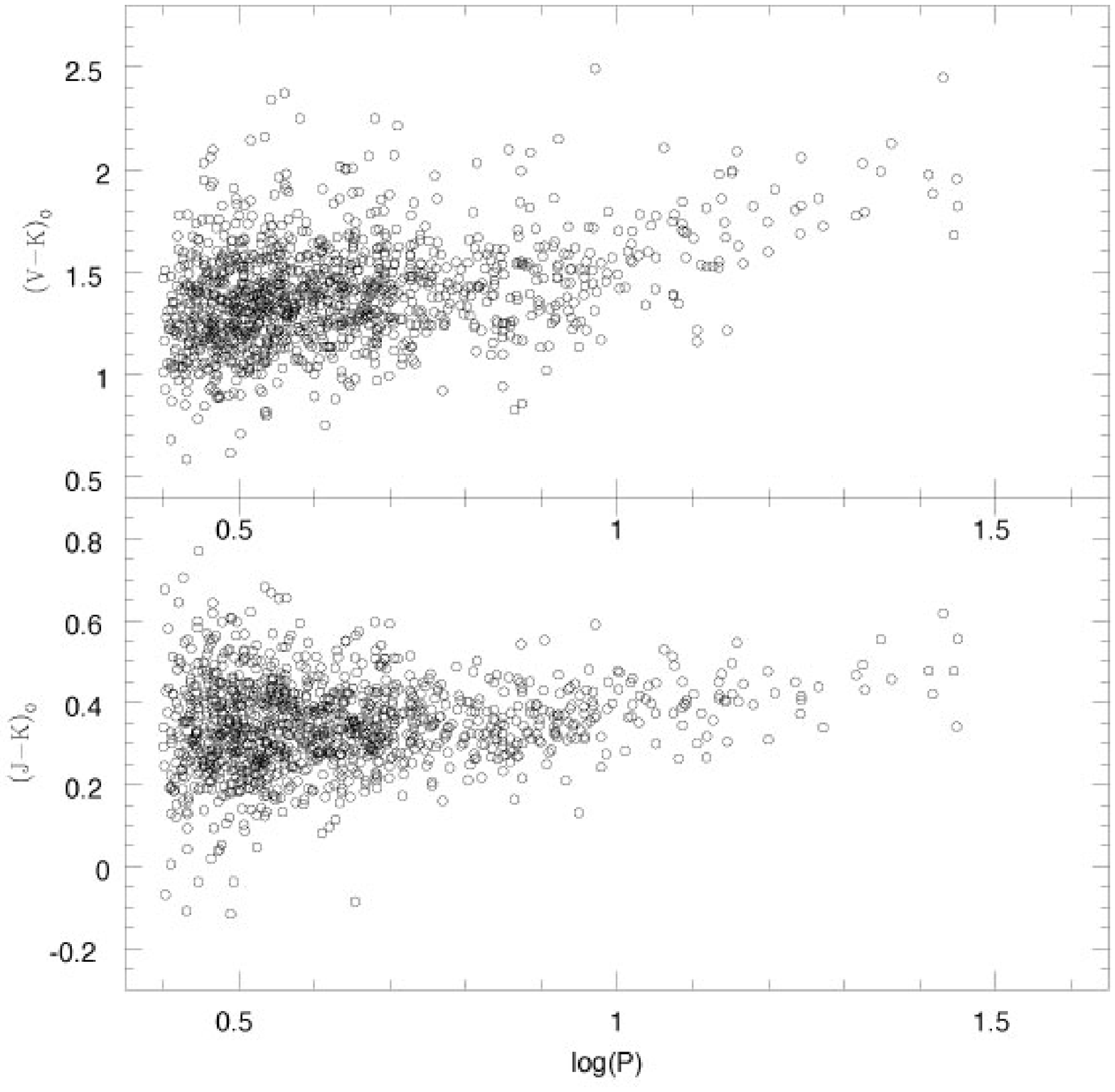}}
       \vspace{0cm}
       \caption{The selected PC relations after extinction correction, constructed with the Sample C data as mentioned in Section 2.}
       \label{figpc}
     \end{figure*}

    The reason that the PL relation is non-linear is because the PC relation for the LMC Cepheids is also non-linear \citep{tam02a,kan04}. This can be clearly seen from upper panels of Figure \ref{figpc} for extinction corrected $(V-R)$ \& $(V-K)$ PC relations. Since both of the PL and PC relations obey the more general PLC relations, the morphology of the PC relation will affect the morphology of the PL relation \citep{mad91}. Recall that the luminosity variation for the Cepheid variables in visual bands is dominated by the temperature variation \citep{cox80}. However, as the waveband moves toward the infrared, the radius variation begin to dominate the variation in the infrared PL relations \citep[e.g.,][]{cox80,mad85}. Since there is no obvious reason that the period-radius relation for LMC Cepheids should be non-linear \citep[see, e.g.,][]{bon98,gie98,gie99}, one would expect that the PL relation is linear toward the infrared. From this study, we see that the PL relation is non-linear at {\it VRJ}- \& {\it H}-band, and marginally linear at {\it K}-band. This suggests that the radius variation may begins to dominate at {\it K}-band and longer wavelengths. This evidence can also be seen from the lower-right panel of Figure \ref{figpc}, where the extinction corrected $(J-K)$ PC relation becomes linear. These findings also imply that the non-linearity of the PL relations is due to the temperature (or colour) variations as a function of period. The detailed discussion and investigation of the physical reason behind the non-linear PL and PC relation is beyond the scope of this paper, and it will be addressed in a future paper.

    Some attempts at using the long period LMC PL relation (from two-line regression) in distance scale studies can be found in \citet{kan03}, \citet{leo03}, \citet{thi03}, \citet{thi04} and \citet{rie05} without any statistical justification. We have shown in Section 3 \& 4 that there is strong statistical evidence that the (optical) LMC PL relation is non-linear, hence the use of long period PL relation is justified. \citet{nge05} has discussed the effect of using the broken LMC PL relations (two-line regression) in distance scale applications: they found that this effect is minimal with a systematic error of $0.03$ mag. or less for the derived distance modulus. Therefore the finding of a non-linear LMC PL relation is more important in terms of the studies of stellar pulsation and evolution to understand the physical reasons that cause the non-linearity than in distance scale applications. However, in the era of precision cosmology, it is important to use the latest understanding of Cepheid PL relations.

     \begin{table}
       \centering
       \caption{The results of the chi-square test, with $H_O=$one-line regression and $\log(P_o)=1.0$. }
       \label{tabch2a}
       \begin{tabular}{lccc} \hline 
         $H_A=$  & \multicolumn{1}{c}{two-lines} & \multicolumn{1}{c}{piecewise} & \multicolumn{1}{c}{quadratic} \\
         Bandpass  & $\chi^2$ & $\chi^2$ & $\chi^2$ \\     
         \hline 
         \multicolumn{4}{c}{Sample A, $N=1330$} \\
         $V\dots$ & 19.47 & 19.41 & 1.282   \\
         $R\dots$ & 21.26 & 21.16 & 1.869   \\
         $J\dots$ & 11.69 & 10.14 & 3.222   \\
         $H\dots$ & 11.14 & 9.687 & 6.063   \\
         $K\dots$ & 3.870 & 3.127 & 0.426   \\
         \multicolumn{4}{c}{Sample B, $N=1322$} \\
         $V\dots$ & 23.35 & 22.99 & 4.235   \\
         $R\dots$ & 25.34 & 24.89 & 5.235   \\
         $J\dots$ & 13.06 & 11.14 & 4.652   \\
         $H\dots$ & 11.91 & 10.23 & 7.276   \\
         $K\dots$ & 4.166 & 3.321 & 0.631   \\
         \multicolumn{4}{c}{Sample C, $N=1216$} \\
         $V\dots$ & 30.85 & 29.60 & 14.38   \\
         $R\dots$ & 33.40 & 31.94 & 16.86   \\
         $J\dots$ & 16.02 & 13.26 & 9.553   \\
         $H\dots$ & 11.52 & 9.778 & 7.527   \\
         $K\dots$ & 4.981 & 3.877 & 1.698   \\
         \hline
       \end{tabular}
     \end{table}

     \begin{table}
       \centering
       \caption{The results of the $F$-test, with $H_O=$one-line regression and $\log(P_o)=1.0$. }
       \label{tabfa}
       \begin{tabular}{lccc} \hline 
         $H_A=$  & \multicolumn{1}{c}{two-lines} & \multicolumn{1}{c}{piecewise} & \multicolumn{1}{c}{quadratic} \\
         Bandpass  & $F$ & $F$ & $F$ \\     
         \hline 
         \multicolumn{4}{c}{Sample A, $N=1330$} \\
         $V\dots$ & 6.367 & 12.71 & 0.832  \\
         $R\dots$ & 7.178 & 14.30 & 1.251  \\
         $J\dots$ & 8.988 & 15.58 & 4.908  \\
         $H\dots$ & 9.539 & 16.57 & 10.32  \\
         $K\dots$ & 2.923 & 4.723 & 0.642  \\
         \multicolumn{4}{c}{Sample B, $N=1322$} \\
         $V\dots$ & 9.428 & 18.58 & 3.383  \\
         $R\dots$ & 10.53 & 20.70 & 4.301  \\
         $J\dots$ & 10.91 & 18.57 & 7.692  \\
         $H\dots$ & 10.68 & 18.33 & 12.98  \\
         $K\dots$ & 3.182 & 5.072 & 0.960  \\
         \multicolumn{4}{c}{Sample C, $N=1216$} \\
         $V\dots$ & 12.32 & 23.64 & 11.37  \\
         $R\dots$ & 13.76 & 26.31 & 13.75  \\
         $J\dots$ & 13.37 & 22.07 & 15.83  \\
         $H\dots$ & 10.30 & 17.46 & 13.40  \\
         $K\dots$ & 3.788 & 5.891 & 2.573  \\
         \hline
       \end{tabular}
     \end{table}

     We now discuss some relevant issues that may affect our results in the following subsections.

\subsection{Issue of outliers}

     Could outliers be one cause for our results? -- KN tested this extensively for both PL and PC relations and concluded that outliers cannot be responsible for the break. Figure \ref{figmulti}, using OGLE data and depicting both PC and PL relations at pulsation phase close to 0.82, is remarkably clear of outliers: the break or non-linearity at a period of 10 days is compelling. The fact that our analysis of MACHO/2MASS data produce results consistent with our previous analysis of the OGLE data set suggests that outliers cannot be responsible for the results presented in this paper. Nevertheless, we adopt a robust regression method which limits the influence of outliers by using Tukey's Bi-weight function as, e.g., given in \citet{pre92} to the {\it V}-band data with two-line vs. one-line regression. The general idea is to suppress the influence of points with large deviation from the expected model. The $F$-test results are: $F_V=5.869$ ($p=0.003$) for Sample A, $F_V=5.924$ ($p=0.003$) for Sample B and $F_V=9.196$ ($p=0.000$) for Sample C, respectively. Minimizing the sum of the absolute deviations rather than the sum of the squares also yields broadly similar results. Another approach is the {\tt RANSAC} algorithm \citep{fis81,sto04} used to detect linear structure in the presence of many outliers. This algorithm tries to find out how much support a given model has among the data. Equation (1) of \citet{sto04} shows how many times we have to repeat the algorithm for a given probability of failure. Given that we have over $\sim1300$ data points this algorithm only makes sense if we have over $\sim200$ outliers and we do not believe this to be the case.  We also tried the sigma-clipping algorithm \citep{uda99} to remove more outliers, and again we found that our results remain unchanged.
       
\begin{figure}
       \hbox{\hspace{0.1cm}\epsfxsize=7.5cm \epsfbox{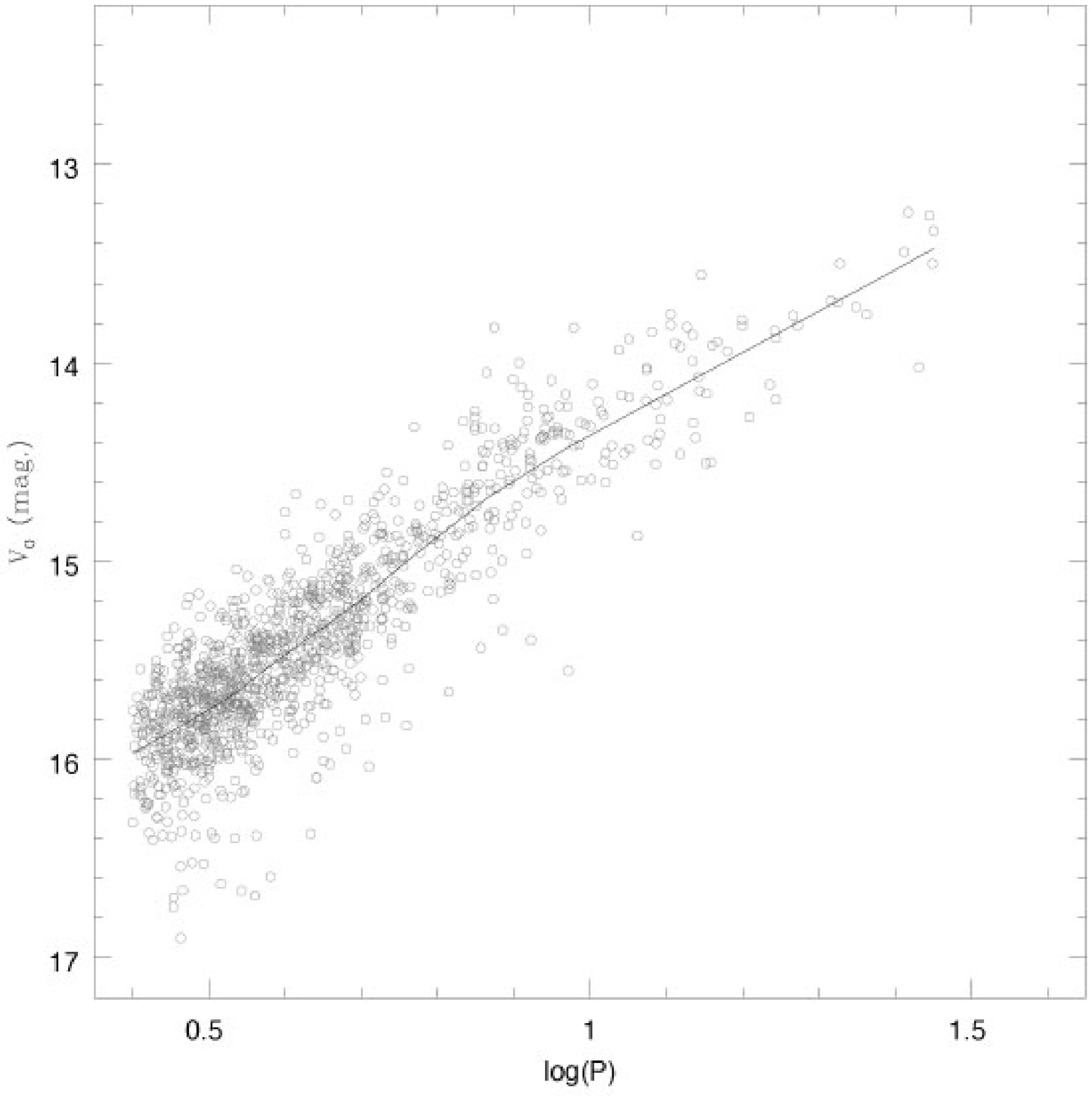}}
       \caption{A robust non-parametric fit (solid line) between $\log (P)$ and the extinction corrected {\it V}-band MACHO data.}
      \label{fignonpar}
     \end{figure}

       All the above methods are parametric and make specific assumptions about the form of the regression surface. The {\tt LOESS} (Local Linear Smoothing) procedure in {\tt SAS} implements a non-parametric method for implementing a non-parametric regression surface \citep{cle79}. If we have $y(i) = g[x(i)] + {\epsilon}(i)$, where $x,y$ are the predictors and independent variables respectively, {\tt LOESS} works by approximating the value, $g$, of the regression function locally, say at $x_0$, by the value of a function in some specified parametric class. This is done by fitting a regression surface within a chosen neighborhood of $x_0$. The fraction of the data points contained in this neighborhood is the smoothing parameter. This procedure is suitable when there are outliers and robust techniques are required. Figure \ref{fignonpar} portrays our results using this method when applied to the extinction corrected MACHO {\it V}-band data in Sample C. The solid line in this figure is the non-parametric fit and the points are the original data. The non-linearity of the fit between $0.8 < \log(P)< 1.1$ is self-evident, and these results are insensitive to the smoothing parameter used. What we take from this is that our results are, to a large extent, independent of fitting method and robust to outliers.
    
\subsection{Issue of long period Cepheids}

     Could the small number/incompleteness at the long period end be responsible to our results? -- If this is the case, then these effects should act in the same way at all phases. Again Figure \ref{figmulti} displays a preliminary result: the reddening corrected PC and PL relations for OGLE data at the pulsation phase of 0.82 in the left and right panels, respectively. If the true underlying PC relation at this phase is indeed linear, then the combination of selection effects/incompleteness/period cuts acts so as to make the longer period Cepheids appear cooler than they really are. From observations, longer period Cepheids obey a flat PC relation \citep{cod47,sim93,kan04,kan04b,kan05}. Figure 2 of KN provides evidence for this for LMC Cepheids with the OGLE data. We have also evaluated the $(V-R)$ colour at maximum light for the MACHO LMC Cepheid sample and also find a flat PC relation for these stars at maximum light (Kanbur et al 2005 -- in preparation). However, if a combination of small number/incompleteness at the long period end/selection effects is responsible for our results, then the true distribution at maximum light must be such that longer period Cepheids get hotter or bluer as the period increases\footnote{Note that Cepheids can generally get cooler as periods increase, due to the morphology of the instability strip.}. There is very credible physics that predicts the true PC relation at maximum light should be flat \citep{sim93,kan04b}. Thus it is our contention that a combination of selection/incompleteness/period cuts cannot be responsible for our results.

       The small number of long period Cepheids is expected as \citet{alc99} and \citet{uda99b} showed that the period distribution for the LMC Cepheids peaked at $\log(P)\sim0.5$ with a long tail extended out to the long period end. This is because in general Cepheids with longer period have higher mass. It is well known that due to the initial mass function ($dN/dM\propto M^{-\alpha}$, where $M$ is the mass of the stars and $\alpha>0$), the number of high mass stars is expected to be less than the low mass stars \citep[see, e.g.,][]{alc99}. In addition, the crossing time (across the instability strip) for the high mass stars is shorter than their low mass counterparts \citep{bon00}. These two effects reduce the number of long period Cepheids in comparison to the number of short period Cepheids. To show our results do not suffer from the missing long period Cepheids with $\log(P)>1.5$, we add extinction corrected {\it V}-band data for the 18 longest period (i.e, $\log[P]>1.5$) Cepheids from \citet{san04} to our sample. We then apply the $F$-test to the combined sample and test whether the data are more consistent with a one-line or two-line regression. The resulting $F$-value is $8.734$, which implies that the LMC {\it V}-band PL relation is still non-linear with the additional data. We also tried to include all 97 Cepheids (with $0.4<\log[P]<2.2$) from \citet{san04} to our sample, and the $F$-test result still supports the non-linear PL relation.

       A more rigorous test, as suggested by an anonymous referee, is to increase the dataset by adding long period Cepheids from a homogeneous dataset that does not belong to the MACHO sample. Such data exist in the literature, for example from \citet{seb02} which we will use in our test. We first match the Cepheid coordinates and periods in our MACHO data and \citet{seb02} sample to identify the common Cepheids in both samples. We find 203 such stars. For these common Cepheids, we calculate the average difference, $\overline{\Delta}$, between the {\it V} and {\it R} mean magnitudes in these two samples, in the sense that $\Delta_V=V_{MACHO}-V_{Sebo}$ and $\Delta_R=R_{MACHO}-R_{Sebo}$. After removing some obvious outliers ($\Delta > 1$), we obtain: $\overline{\Delta_V}=0.041$ (with std $=0.121$) and $\overline{\Delta_R}=0.039$ (with std $=0.120$). This step is to ensure the \citet{seb02} photometric system is the same as MACHO photometric system by adding $\overline{\Delta}$ to the \citet{seb02} mean magnitudes. In addition, we also found that there is no period dependency in this transformation: $\Delta_V= 0.037(\pm0.034) \log (P) + 0.013(\pm0.027)$, $\Delta_R=-0.028(\pm0.034) \log (P) + 0.060(\pm0.027)$.  If there are any dependencies with period in such a transformation, it could be that this has influenced our results. We see clearly that this is not the case. Similarly, the average difference in {\it JHK} mean magnitudes between the \citet{per04} sample and the MACHO/2MASS sample is negligible, and more importantly, independent of periods (see Section 2 for details).
       
       These photometric transformations (albeit very close to zero) can now be used to transform the \citet{seb02} data to the same photometric
       system as the MACHO {\it V}- and {\it R}-band data used in this study, and to transform the MACHO/2MASS {\it JHK} data to the same system as the \citet{per04} data. This procedure provides a consistent way of increasing our long period sample by identifying those stars not in the MACHO/2MASS sample but which
       are in either the \citet{seb02} or \citet{per04} samples. We now describe our results when this is carried out. 

       The long period ($\log[P]>1.0$) Cepheids in the \citet{seb02} sample that are not in the MACHO sample are readily identified. We found 49 of them\footnote{Some OGLE Cepheids identified by \citet{seb02} are not fundamental mode Cepheids and we exclude them in the table.} and they are listed in Table \ref{long}. In this table, we give the original {\it V} and {\it R} mean magnitudes from \citet{seb02}, as well as the number of observed data points ($m_{V,R}$) for each Cepheids in either bands. The periods are adopted form either \citet{per04} or the OGLE database \citep{uda99b}. The table also lists the {\it JHK} mean magnitudes and the $E(B-V$) values for these Cepheids that are compiled from \citet{per04} and \citep{uda99b}, or we set $E(B-V)=0.10$ if no entries are given in \citet{per04}. Since the \citet{seb02} database lacks Cepheids between $\log(P)\sim1.55$ and $\log(P)\sim1.95$, we include 8 Cepheids in this period range (with the available {\it JHK} mean magnitudes) from \citet{per04} in Table \ref{long} for completeness. We did not attempt to add the {\it V} and {\it R} mean magnitude for these 8 Cepheids from the literature, as in \citet{san04}, in order to keep our dataset as homogeneous and consistent as possible.

     \begin{table*}
       \centering
       \caption{Additional long period Cepheids from \citet{seb02} and \citet{per04}.}
       \label{long}
       \begin{tabular}{lccccccccc} \hline 
        Name  &   $\log(P)$  & $m_V$  & $m_R$ &  $V$  & $R$  & $J$ & $H$ & $K$ & $E(B-V)$ \\
         \hline 
         Ogle100794 & 1.0298 & 17 & 17 & 14.78 & 14.30 & $\cdots$ & $\cdots$ & $\cdots$ & 0.138 \\
         Ogle50505  & 1.0356 & 14 & 13 & 14.69 & 14.23 & $\cdots$ & $\cdots$ & $\cdots$ & 0.127 \\
         HV2432     & 1.0384 & 7  & 5  & 14.24 & 13.96 & 12.986 & 12.653 & 12.568 & 0.100 \\
         Ogle96614  & 1.0862 & 20 & 19 & 14.58 & 14.10 & $\cdots$ & $\cdots$ & $\cdots$ & 0.138 \\
         Ogle117525 & 1.1016 & 18 & 16 & 14.43 & 13.92 & $\cdots$ & $\cdots$ & $\cdots$ & 0.128 \\
         Ogle173734 & 1.1045 & 6  & 4  & 14.72 & 14.20 & $\cdots$ & $\cdots$ & $\cdots$ & 0.135 \\
         HV2527     & 1.1122 & 33 & 32 & 14.62 & 14.16 & 13.084 & 12.684 & 12.570 & 0.070 \\
         HV2260     & 1.1136 & 16 & 14 & 14.85 & 14.35 & 13.263 & 12.860 & 12.736 & 0.130 \\
         Ogle270379 & 1.1233 & 11 & 10 & 14.38 & 13.85 & $\cdots$ & $\cdots$ & $\cdots$ & 0.138 \\
         HV2579     & 1.1281 & 10 & 7  & 14.04 & 13.63 & 12.688 & 12.334 & 12.236 & 0.100 \\
         HV955      & 1.1377 & 8  & 7  & 14.02 & 13.62 & 12.686 & 12.342 & 12.251 & 0.058 \\
         HV2538     & 1.1420 & 6  & 3  & 14.50 & 14.05 & 12.864 & 12.452 & 12.343 & 0.100 \\
         HV2463     & 1.1450 & 10 & 8  & 14.22 & 13.81 & 12.748 & 12.381 & 12.295 & 0.100 \\
         HV5655     & 1.1526 & 16 & 11 & 14.52 & 14.06 & 12.971 & 12.553 & 12.443 & 0.100 \\
         Ogle167787 & 1.1610 & 11 & 6  & 14.09 & 13.62 & $\cdots$ & $\cdots$ & $\cdots$ & 0.145 \\
         HV12471    & 1.2001 & 9  & 4  & 14.73 & 13.95 & 12.903 & 12.453 & 12.308 & 0.058 \\ 
         Ogle55470  & 1.2039 &  9 &  6 & 14.65 & 14.08 & $\cdots$ & $\cdots$ & $\cdots$ & 0.121 \\
         Ogle182466 & 1.2039 & -- & -- & 14.31 & 14.08 & $\cdots$ & $\cdots$ & $\cdots$ & 0.127 \\
         HV2549     & 1.2094 & 9  & 8  & 13.67 & 13.36 & 12.422 & 12.078 & 11.982 & 0.058 \\
         Ogle160625 & 1.2358 & 19 & 14 & 13.78 & 13.34 & $\cdots$ & $\cdots$ & $\cdots$ & 0.138 \\
         HV2261     & 1.2370 & 11 & 8  & 13.26 & 12.90 & $\cdots$ & $\cdots$ & $\cdots$ & 0.100 \\
         HV2580     & 1.2285 & 17 & 12 & 14.01 & 13.53 & 12.489 & 12.102 & 11.999 & 0.090 \\
         HV2836     & 1.2437 & 13 & 9  & 14.63 & 14.07 & 12.747 & 12.276 & 12.127 & 0.180 \\
         Ogle250872 & 1.2709 & 10 & 11 & 13.97 & 13.52 & $\cdots$ & $\cdots$ & $\cdots$ & 0.152 \\
         U11        & 1.3026 & 9  & 6  & 13.66 & 13.23 & 12.383 & 11.972 & 11.860 & 0.100 \\
         Ogle148920 & 1.3160 & 21 & 18 & 13.60 & 13.08 & $\cdots$ & $\cdots$ & $\cdots$ & 0.147 \\
         Ogle88441  & 1.3276 & 4  & 4  & 13.31 & 13.00 & $\cdots$ & $\cdots$ & $\cdots$ & 0.142 \\
         Ogle286532$^{\mathrm a}$ & 1.3347 & 12 & 12 & 14.01 & 13.44 & $\cdots$ & $\cdots$ & $\cdots$ & 0.138 \\
         Ogle109640 & 1.3486 & 18 & 16 & 14.00 & 13.51 & $\cdots$ & $\cdots$ & $\cdots$ & 0.138 \\
         U1         & 1.3533 & 9  & 5  & 14.09 & 13.62 & 12.350 & 11.916 & 11.790 & 0.100 \\
         HV876      & 1.3561 & 10 & 7  & 13.53 & 13.32 & 12.144 & 11.766 & 11.665 & 0.100 \\
         HV878      & 1.3673 & 33 & 27 & 13.49 & 13.07 & 12.097 & 11.735 & 11.634 & 0.058 \\
         HV938      & 1.3724 & 5  & 5  & 13.31 & 12.91 & $\cdots$ & $\cdots$ & $\cdots$ & 0.100 \\
         HV6098     & 1.3845 & 8  & 6  & 12.95 & 12.56 & 11.733 & 11.405 & 11.317 & 0.100 \\
         HV902      & 1.4209 & 6  & 4  & 13.25 & 12.70 & 11.833 & 11.469 & 11.378 & 0.070 \\
         HV1023     & 1.4239 & 22 & 19 & 13.77 & 13.20 & 12.041 & 11.618 & 11.498 & 0.070 \\
         Ogle162232 & 1.4828 & 16 & 11 & 13.48 & 12.88 & $\cdots$ & $\cdots$ & $\cdots$ & 0.152 \\
         HV872      & 1.4750 & 11 & 9  & 13.69 & 13.15 & 11.976 & 11.540 & 11.422 & 0.100 \\
         HV875      & 1.4822 & 16 & 16 & 12.96 & 12.58 & 11.620 & 11.285 & 11.192 & 0.100 \\
         Ogle228645 & 1.4921 & 11 & 7  & 13.26 & 12.83 & $\cdots$ & $\cdots$ & $\cdots$ & 0.129 \\
         HV882      & 1.5027 & 18 & 15 & 13.35 & 12.90 & 11.714 & 11.328 & 11.202 & 0.070 \\
         HV5761     & 1.5046 & 5  & 3  & 13.14 & 12.38 & $\cdots$ & $\cdots$ & $\cdots$ & 0.100 \\
         HV873      & 1.5359 & 16 & 14 & 13.06 & 12.58 & 11.490 & 11.109 & 10.998 & 0.130 \\
         HV881      & 1.5534 & 16 & 15 & 13.09 & 12.61 & 11.527 & 11.142 & 11.032 & 0.030 \\
         HV2294     & 1.5626 & -- & -- & $\cdots$ & $\cdots$ & 11.237 & 10.868 & 10.770 & 0.070 \\
         HV879      & 1.5656 & -- & -- & 13.64 & 13.01 & 11.590 & 11.157 & 11.031 & 0.060 \\
         HV909      & 1.5749 & -- & -- & 13.03 & 12.54 & 11.339 & 10.984 & 10.884 & 0.058 \\
         HV2257     & 1.5943 & 9  & 8  & 13.05 & 12.53 & 11.355 & 10.944 & 10.833 & 0.060 \\
         HV2338     & 1.6248 & -- & -- & $\cdots$ & $\cdots$ & 11.185 & 10.798 & 10.690 & 0.040 \\
         HV877      & 1.6548 & -- & -- & $\cdots$ & $\cdots$ & 11.438 & 10.962 & 10.835 & 0.100 \\
         HV900      & 1.6771 & -- & -- & $\cdots$ & $\cdots$ & 11.156 & 10.751 & 10.632 & 0.058 \\
         HV953      & 1.6802 & -- & -- & $\cdots$ & $\cdots$ & 10.788 & 10.423 & 10.317 & 0.070 \\
         HV2827     & 1.8969 & -- & -- & $\cdots$ & $\cdots$ & 10.428 &  9.976 &  9.851 & 0.080 \\
         HV5497     & 1.9965 & 9  & 6  & 11.93 & 11.38 & 10.027 & 9.602  & 9.466  & 0.095 \\
         HV2883$^{\mathrm b}$     & 2.0398 & -- & -- & $\cdots$ & $\cdots$ & 10.651 & 10.237 & 10.098 & 0.100 \\
         HV2447     & 2.0763 & -- & -- & $\cdots$ & $\cdots$ & 10.106 &  9.704 &  9.565 & 0.100 \\
         HV883      & 2.1268 & 8  & 6  & 12.00 & 11.35 & 10.302 & 9.919  & 9.752  & 0.100 \\
         \hline
       \end{tabular}
       \begin{list}{}{}
       \item $^{\mathrm a}$This Cepheids has relatively low amplitude \citep{kan05}.
       \item $^{\mathrm b}$This Cepheids has relatively high amplitude \citep{kan05}.
       \end{list}
     \end{table*}

       The data in Table \ref{long} are then added to our MACHO/2MASS sample, after the appropriate photometric transformation. The $F$-tests are applied to this extended sample for $H_A=$ two-lines regression (with a break at 10 days) using the following selection criteria:

       \begin{enumerate}
       \item Include all available long period Cepheids in Table \ref{long}, appropriate to the bandpass, to our MACHO/2MASS sample.
       \item Eliminate Cepheids without {\it JHK} mean magnitudes since not all of the Cepheids in Table \ref{long} have the {\it JHK} means.
       \item Eliminate Cepheids with $m_{V,R}\leq 10$ as the mean magnitudes derived from these sparsely sampled light curves may not be accurate.
       \item Combine step (ii) \& (iii) above (item 5 in Table \ref{tabflong}) in order to have consistent number of Cepheids in all five bands. 
       \end{enumerate}

     \begin{table*}
       \centering
       \caption{$F$-test results with additional long period Cepheids, where $N_{long}^{tot}$ is the total number of long period Cepheids (MACHO + additional Cepheids from Table \ref{long}). }
       \label{tabflong}
       \begin{tabular}{lcccccccccc} \hline 
         Bandpass  & \multicolumn{2}{c}{$V$} & \multicolumn{2}{c}{$R$} & \multicolumn{2}{c}{$J$} & \multicolumn{2}{c}{$H$} & \multicolumn{2}{c}{$K$} \\
         Criteria  & $N_{long}^{tot}$ & $F$ & $N_{long}^{tot}$ & $F$ & $N_{long}^{tot}$ & $F$ & $N_{long}^{tot}$ & $F$ & $N_{long}^{tot}$ & $F$ \\     
         \hline 
         1. use all available long period Cepheids   & 118 & 7.872 & 118 & 8.640 & 106 & 12.33 & 106 & 10.61 & 106 & 2.942  \\
         2. exclude Cepheids without {\it JHK} means &  98 & 10.96 &  98 & 12.45 &  -- &   --  &  -- &   --  &  -- &    -- \\
         3. Cepheids with $m_V>10$                   &  94 & 8.748 &  -- &  --   &  81 & 13.71 &  81 & 10.29 &  81 & 3.849  \\ 
         4. Cepheids with $m_R>10$                   &  -- &  --   &  89 & 10.32 &  79 & 12.70 &  79 & 9.603 &  79 & 3.681  \\
         5. combine item (2), (3) and (4)            &  79 & 10.86 &  79 & 12.38 &  79 & 12.70 &  79 & 9.603 &  79 & 3.681 \\
         \hline
       \end{tabular}
     \end{table*}

       \ni The results from these different selection criteria are presented in Table \ref{tabflong}. Note that \citet{per04} did not use Cepheids with $\log(P)>2.0$. These Cepheids are eliminated in step (iv) above (item 5 in Table \ref{tabflong}). The linearity of the PL relation in each band again can be tested by comparing the $F$-values in this table with Table \ref{tabp}. As can be seen from the table, including the additional long period Cepheids does not alter the main conclusion we have: the {\it VRJH}-band PL relation is non-linear and the {\it K}-band PL relation is (marginally) linear. Since there are no measurement errors given in \citet{seb02} for the {\it V} and {\it R} mean magnitudes, the results in Table \ref{tabflong} are derived with unweighted regression (i.e. $w=1$) in all bands. The $F$-values from using the weighted regressions to all bands, with assumed some values for the errors in the {\it V}- and {\it R}-band means (and the $E[B-V]$, the errors for {\it JHK} means are given in \citealt{per04}), are very similar to Table \ref{tabflong}. One exception is the $F$-value for the {\it K}-band when using all available long period Cepheids, the $F$-value goes down from $2.942$ to $1.532$, which is fully consistent with a one-line regression. In this table, except for the {\it K}-band, the lowest significant $F$-value is 7.872. With the degrees of freedom appropriate for the data in our study, this has a $p$-value less than $10^{-3}$ (recall that $p<10^{-3}$ for $F>7$). This means that assuming the null hypothesis of a single regression line being sufficient. The chance of getting a value of the $F$ statistic as high or greater than that actually observed (7.872) is less than $10^{-3}$.

\subsection{Issue of reddening/extinction}

      Could reddening and extinction errors produce our results? -- Figure \ref{figpc} displays extinction corrected PC relations, at mean light, obtained from the MACHO+2MASS data for a variety of colours: $(V-R)_0,\ (V-J)_0,\ (V-K)_0$ \& $(J-K)_0$. This figure and the PC relations presented in KN and \citet{san04} imply that in order to make a linear PC relation, extinction errors would have to be of the order of $\sim0.05$-$0.2$ mag. for the longer period Cepheids. If this were indeed the case, the same error would apply to PC relations at maximum light. Again, longer period Cepheids obey a flat PC relation at maximum light (KN, SKM, Code 1947), an observation for which there is a sound theoretical argument (see the reference given in Section 5.2). If these longer period Cepheids suffered an extinction error of $\sim0.05$-$0.2$ mag., then the true underlying PC relation at maximum light would be such that as the period increases the colour becomes bluer (that is the temperature is hotter) at maximum light. It is hard to incorporate this in the theoretical framework of \citet{sim93}, since this paper suggested that for very long period Cepheids ($\log[P] > 1.5$ roughly), the PC relation at maximum light then starts to have a slope such that Cepheids become redder as the period gets longer. This is in the opposite way to what would need to be true were the PC relation to be linear at mean light. Further, if reddening/extinction errors are indeed making the PC/PL relation appear to be non-linear then the extinction/reddening from two independent methods, as used by the OGLE team and that used in this study, would both need to contain errors of the order of $\sim0.05$-$0.2$ mag. for the longer period Cepheids.
  
       Although longer period Cepheids are generally younger and as such may be surrounded by dust shells which could increase extinction with period and possibly produce a non-linear PC relation as is observed. We argue again that our knowledge of the properties of Cepheids at maximum light as presented above argues against this. Further such an effect would be gradual, however the observed PC relation is sharply non-linear as is evident in the left panel of Figure \ref{figmulti}. The use of two extinction maps, one for the OGLE data and another one in this study, to correct for the extinction/reddening that produce similar results also suggest the effect of the extinction is minimal. 

\begin{figure*}
       \vspace{0cm}
       \hbox{\hspace{0.2cm}\epsfxsize=8.5cm \epsfbox{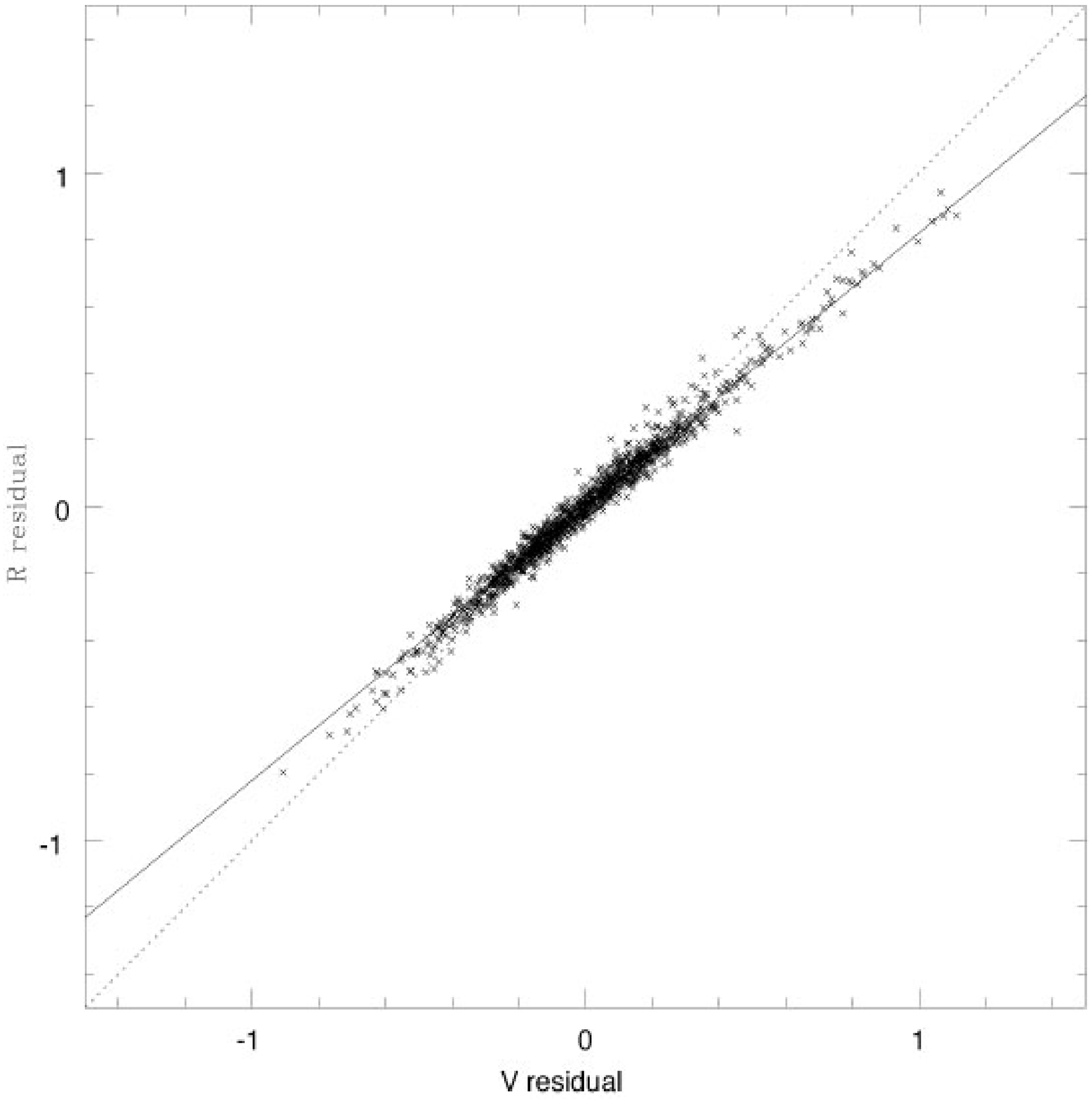}
         \epsfxsize=8.5cm \epsfbox{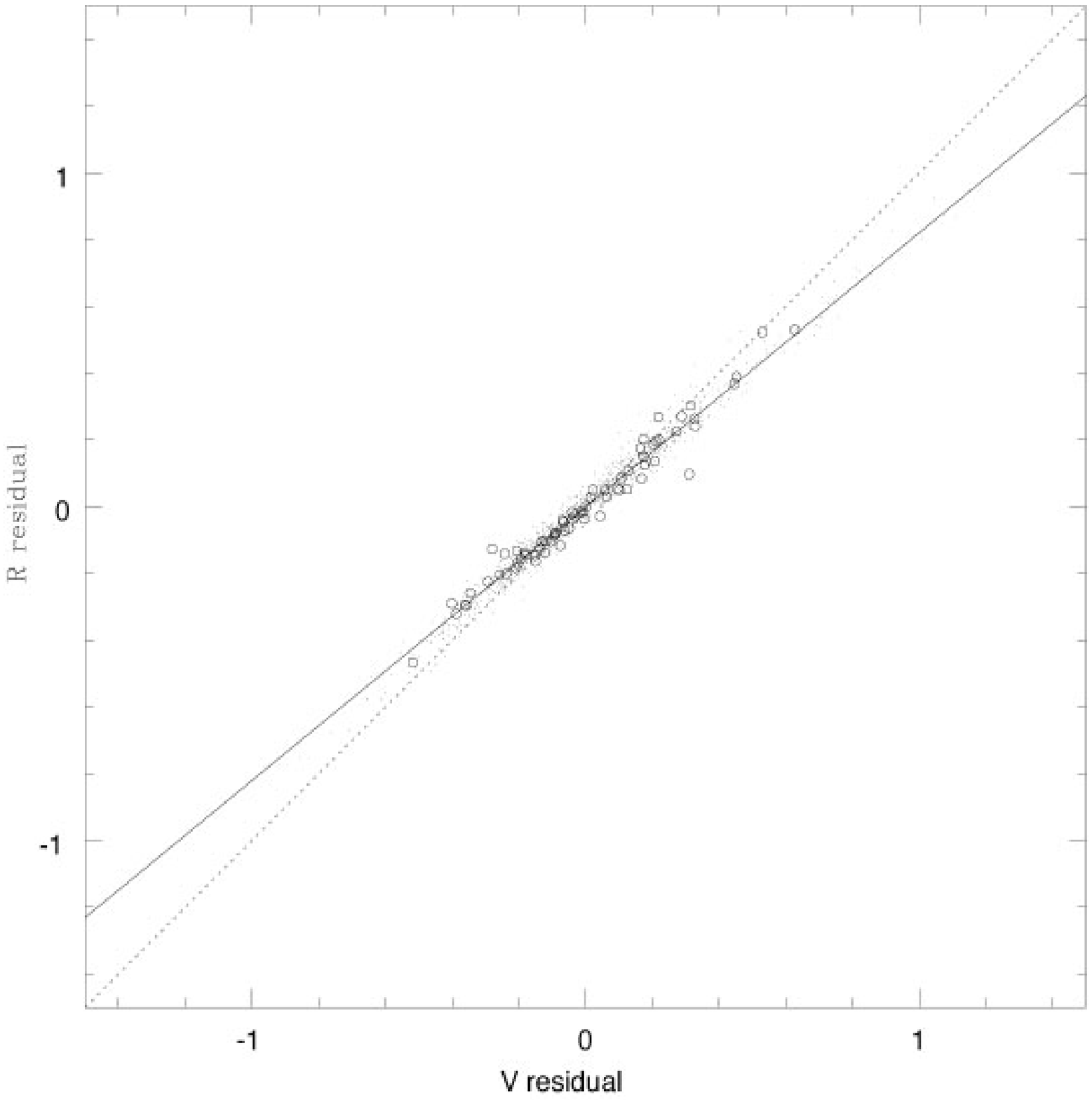}}
       \vspace{0cm}
       \caption{The residual plots of the {\it V}- and {\it R}-band PL relations from one-line regression (left panel) and two-line regression with a break at 10 days (right panel). In the right panel, the dots and open circles represent the residuals from the short (less than 10 days) and long period Cepheids, respectively. The solid and dashed lines represent the reddening lines (defined by our reddening vector) and the expected residuals due to depth dispersion (diagonal lines), respectively.}
       \label{fig7}
     \end{figure*}

       The anonymous referee has also suggested the following to test to examine if our extinction corrections do indeed represent individual corrections: plotting the residuals from the {\it V}- and {\it R}-band PL relations against each other \citep[see, e.g][]{sas97,bea01}. The idea is that if the residuals lie along the reddening line, then the reddening corrections do not represent individual corrections. We carried out such a test with Sample C using the PL relations from a one-line and two-line regression. The results are presented in Figure \ref{fig7}. From the figure it can be seen that the dispersion is reduced when using the one-line regressions as compared to the two-line regression. The points lie along the reddening line. 

       We, however, argue that this test is not suitable for testing the extinction corrections for the following reasons. First of all, besides the statistical fluctuations, the residuals from the PL relation are expected to originate from three sources \citep{sas97}: (a) depth dispersion/effects (b) extinction, and (c) intrinsic dispersion due to the width of the instability strip. All three sources will tend to make the residuals be correlated in the {\it V}- and {\it R}-bands, as seen in Figure \ref{fig7}. We ignore the discussion of the depth effect (represented as dashed lines in Figure \ref{fig7}) because we assume the depth effect is negligible for the LMC Cepheids. For the source from extinction, recall that the the extinction correction for a Cepheid is calculated using $A_{\lambda}=R_{\lambda}E(B-V)$, where the $E(B-V)$ value is either the mean value of the LMC or an estimated value for individual Cepheids. Hence a correlation of the residuals is naturally introduced (represented as solid lines in Figure \ref{fig7}) for the reddening vector used in this paper (Section 2), because $R_{\lambda}/R_V\simeq$ constant. Even for a hypothetical group of Cepheids that are free from extinction and depth dispersion, the residuals of the {\it V}- and {\it R}-band PL relations will still be correlated. This is due to the existence of the instability strip and the PLC relation \citep{san58,mad91}. Therefore mapping the Cepheids from the instability strip (along the constant-period line) to the {\it V}- and {\it R}-band PL relations will cause the residuals to be unavoidably correlated. Unfortunately, this line of residuals is very close to the reddening line in the residual plots (not shown in Figure \ref{fig7}, see also \citealt{sas97}). This makes it hard to say whether the correlation of the residuals seen in Figure \ref{fig7} is caused by the extinction or the width of the instability strip\footnote{The spread of the residuals ($\sim \pm1.0$ mag.) is comparable to the findings of \citet{sas97} and \citet{bea01}. This is not a surprise because:(a) Cepheids in the observed instability strip are roughly normally distributed (with $FWHM\sim \sigma_{IS}$ and dominate the overall distribution); and (b) the measurement and extinction errors are also roughly normally distributed. Hence the overall distribution is also normally distributed with tails extended to $\sim \pm1.0$ mag. at both ends.}.

       In this paper we have tried our best to correct for the extinction of the LMC Cepheids. Based on the above arguments we believe that extinction errors are not the physical cause for the observed non-linear LMC PL relations. Other physical reasons, such as the pulsational properties \citep{bon99,cap00} and/or the internal structure \citep{sim93,kan04b,kan05} of the Cepheid variables (we refer to these as ``internal reasons'' as opposed to the ``external reasons'' like dust extinction) may be responsible for the observed non-linear PL relations. We have provided compelling evidence that the (extinction corrected) Cepheid PL relation in the LMC is non-linear around a period of 10 days in the optical {\it V}- \& {\it R}-bands. Assuming the random phase corrections methods used in this study, we also find the PL relation in the {\it J}- \& {\it H}-band to be non-linear around this period but that the {\it K}-band relation is marginally linear. An investigation into the physics behind this is left for future studies.

\section*{acknowledgements}

The authors would like to thank the assistance from D. Zaritsky and J. Harris regarding the usage of LMC extinction map. SN's \& KC's works were performed under the auspices of the U.S. Department of Energy, National Nuclear Security Administration by the University of California, Lawrence Livermore National Laboratory under contract No. W-7405-Eng-48. This publication makes use of data products from the Two Micron All Sky survey (2MASS), which is a joint project of the University of Massachusetts and the Infrared Processing and Analysis Center, funded by the National Aeronautics and Space Administration and the National Science Foundation. The authors would also like to thank an anonymous referee for several comments which helped to improve the manuscript.


\begin{thebibliography}{}
\bibitem[\protect\citeauthoryear{Andreasen}{1988}]{and88} Andreasen, G., 1988, A\&A, 196, 159
\bibitem[\protect\citeauthoryear{Alcock et al.}{1999}]{alc99} Alcock, C., Allsman, R., Alves, D., Axelrod, T., Becker, A., Bennett, D., Bersier, D., Cook, K., et al., 1999, AJ, 117, 920
\bibitem[\protect\citeauthoryear{Alcock et al.}{2000}]{alc00} Alcock, C., Allsman, R., Alves, D., Axelrod, T., Becker, A., Bennett, D., Cook, K., Dalal, N., et al., 2000, ApJ, 542, 281
\bibitem[\protect\citeauthoryear{Baraffe et al.}{1998}]{bar98} Baraffe, I., Alibert, Y., M\'{e}ra, D., Chabrier, G. \& Beaulieu, J.-P., 1998, ApJ, 499, L205 
\bibitem[\protect\citeauthoryear{Baraffe \& Alibert}{2001}]{bar01} Baraffe, I. \& Alibert, Y., 2001, A\&A, 371, 592
\bibitem[\protect\citeauthoryear{Bauer et al.}{1999}]{bau99} Bauer, F., Afonso, C., Albert, J. N., Alard, C., Andersen, J., Ansari, R., Aubourg, E., Bareyre, P., et al., 1999, A\&A, 348, 175
\bibitem[\protect\citeauthoryear{Beaulieu et al.}{2001}]{bea01} Beaulieu, J.-P., Buchler, R. \& Koll\'{a}th, Z., 2001, A\&A, 373, 164
\bibitem[\protect\citeauthoryear{Bersier et al.}{1994}]{ber94} Bersier, D., Burki, G. \& Burnet, M., 1994, A\&AS, 108, 9
\bibitem[\protect\citeauthoryear{Bono et al.}{1998}]{bon98} Bono, G., Caputo, F. \& Marconi, M., 1998, ApJ, 497, L43
\bibitem[\protect\citeauthoryear{Bono et al.}{1999}]{bon99} Bono, G., Caputo, F., Castellani, V. \& Marconi, M., 1999, ApJ, 512, 711
\bibitem[\protect\citeauthoryear{Bono et al.}{2000a}]{bon00} Bono, G., Caputo, F., Cassisi, S., Marconi, M., Piersanti, L. \& Tornamb\'{e}, A., 2000a, ApJ, 543, 955
\bibitem[\protect\citeauthoryear{Bono et al.}{2000b}]{bon00b} Bono, G. Marconi, M. \& Stellingwerf, R., 2000b, A\&A, 360, 245
\bibitem[\protect\citeauthoryear{Caputo et al.}{1999}]{cap99} Caputo, F., Marconi, M. \& Ripepi, V., 1999, ApJ, 525, 784 
\bibitem[\protect\citeauthoryear{Caputo et al.}{2000}]{cap00} Caputo, F., Marconi, M. \& Musella, I., 2000, A\&A, 354, 610 
\bibitem[\protect\citeauthoryear{Coulson et al.}{1985}]{cou85} Coulson, I., Caldwell, J. \& Gieren, W., 1985, ApJS, 57, 595
\bibitem[\protect\citeauthoryear{Cleveland}{1979}]{cle79} Cleveland, W. S., 1979, Journal of the American Statistical Association, 74, 829
\bibitem[\protect\citeauthoryear{Code}{1947}]{cod47} Code, A. D., 1947, ApJ, 106, 309
\bibitem[\protect\citeauthoryear{Cox}{1980}]{cox80} Cox, J., 1980, {\it Theory of Stellar Pulsation}, Princeton University Press, $1^{st}$ Ed.
\bibitem[\protect\citeauthoryear{Feigelson \& Babu}{1992}]{fei92} Feigelson, E. \& Babu, G. J., 1992, ApJ, 397, 55
\bibitem[\protect\citeauthoryear{Fernie}{1967}]{fer67} Fernie, J. D., 1967, AJ, 72, 1327
\bibitem[\protect\citeauthoryear{Fernie}{1969}]{fer69} Fernie, J. D., 1969, PASP, 81, 707
\bibitem[\protect\citeauthoryear{Fisher \& Bolles}{1981}]{fis81} Fisher, M. A., Bolles, R. C., 1981, Comm. of the ACM., 24, 381
\bibitem[\protect\citeauthoryear{Freedman et al.}{2001}]{fre01} Freedman, W., Madore, B., Gibson, B., Ferrarese, L, Kelson, D., Sakai, S., Mould, J., Kennicutt, R., et al., 2001, ApJ, 553, 47
\bibitem[\protect\citeauthoryear{Gieren et al.}{1998}]{gie98} Gieren, W., Fouqu\'{e}, P. \& G\'{o}mez, M., 1998, ApJ, 496, 17
\bibitem[\protect\citeauthoryear{Gieren et al.}{1999}]{gie99} Gieren, W., Moffett, T. \& Barnes, T., 1999, ApJ, 512, 553
\bibitem[\protect\citeauthoryear{Gieren et al.}{2005}]{gie05} Gieren, W., Storm, J., Barnes, T., Fouqu\'{e}, P., Pietrzy\'{n}ski, G. \& Kienzle, F., 2005, ApJ, 627, 224
\bibitem[\protect\citeauthoryear{Isobe et al.}{1990}]{iso90} Isobe, T., Feigelson, E., Akritas, M. \& Babu, G. J., 1990, ApJ, 364, 104
\bibitem[\protect\citeauthoryear{Kanbur et al.}{2002}]{kan02} Kanbur, S., Iono, D., Tanvir, N. \& Hendry, M., 2002, MNRAS, 329, 126 
\bibitem[\protect\citeauthoryear{Kanbur et al.}{2003}]{kan03} Kanbur, S., Ngeow, C., Nikolaev, S., Tanvir, N. \& Hendry, M., 2003, A\&A, 411, 361
\bibitem[\protect\citeauthoryear{Kanbur \& Ngeow}{2004}]{kan04} Kanbur, S. \& Ngeow, C., 2004, MNRAS, 350, 962 (KN)
\bibitem[\protect\citeauthoryear{Kanbur et al.}{2004}]{kan04b} Kanbur, S., Ngeow, C. \& Buchler, R., 2004, MNRAS, 354, 212
\bibitem[\protect\citeauthoryear{Kanbur \& Ngeow}{2005}]{kan05} Kanbur, S. \& Ngeow, C., 2005, MNRAS submitted 
\bibitem[\protect\citeauthoryear{Laney \& Stobie}{1994}]{lan94} Laney, C. \& Stobie, R., 1994, MNRAS, 266, 441 
\bibitem[\protect\citeauthoryear{Leonard et al.}{2003}]{leo03} Leonard, D., Kanbur, S., Ngeow, C. \& Tanvir, N., 2003, ApJ, 594, 247
\bibitem[\protect\citeauthoryear{Madore}{1985}]{mad85} Madore, B., 1985, ApJ, 298, 340
\bibitem[\protect\citeauthoryear{Madore \& Freedman}{1991}]{mad91} Madore, B. \& Freedman, W., 1991, PASP, 103, 933
\bibitem[\protect\citeauthoryear{Marconi et al.}{2005}]{mar05} Marconi, M., Musella, I. \& Fiorentino, G., 2005, ApJ accepted (astro-ph/0506207)
\bibitem[\protect\citeauthoryear{Moffett et al.}{1998}]{mof98} Moffett, T., Gieren, W., Barnes, T. \& G\'{o}mez, M., 1998, ApJS, 117, 135
\bibitem[\protect\citeauthoryear{Ngeow et al.}{2003}]{nge03} Ngeow, C., Kanbur, S., Nikolaev, S., Tanvir, N. \& Hendry, M., 2003, ApJ, 586, 959
\bibitem[\protect\citeauthoryear{Ngeow \& Kanbur}{2005}]{nge05} Ngeow, C.\& Kanbur, S., 2005, MNRAS, 360, 1033
\bibitem[\protect\citeauthoryear{Nikolaev et al.}{2004}]{nik04} Nikolaev, S., Drake, A., Keller, S., Cook, K., Dalal, N., Griest, K., Welch, D. \& Kanbur, S., 2004, ApJ, 601, 260
\bibitem[\protect\citeauthoryear{Persson et al.}{2004}]{per04} Persson, S., Madore, B., Krzemi\'{n}ski, W., Freedman., W., Roth, M. \& Murphy, D., 2004, AJ, 128, 2239  
\bibitem[\protect\citeauthoryear{Press et al.}{1992}]{pre92} Press, W., Teukolsky, S., Vetterling, W. \& Flannery, B., 1992, \textit{Numerical Recipes in C}, Cambridge University Press, $2^{\mathrm{nd}}$ ed.
\bibitem[\protect\citeauthoryear{Riess et al.}{2005}]{rie05} Riess, A., Li, W., Stetson, P., Filippenko, A., Jha, S., Kirshner, R., Challis, P., Garnavich, P. \& Chornock, R., 2005, ApJ, 627, 579
\bibitem[\protect\citeauthoryear{Saha et al.}{2001}]{sah01} Saha, A., Sandage, A., Tammann, G., Dolphin, A., Christensen, J., Panagia, N. \& Macchetto, F., 2001, ApJ, 562, 314
\bibitem[\protect\citeauthoryear{Sandage}{1958}]{san58} Sandage, A., 1958, ApJ,127, 513
\bibitem[\protect\citeauthoryear{Sandage \& Tammann}{1968}]{san68} Sandage, A. \& Tammann, G. A., 1968, ApJ, 151, 531
\bibitem[\protect\citeauthoryear{Sandage et al.}{2004}]{san04} Sandage, A., Tammann, G. A. \& Reindl, B., 2004, A\&A, 424, 43
\bibitem[\protect\citeauthoryear{Sasselov et al.}{1997}]{sas97} Sasselov, D., Beaulieu, J.-P., Renault, C., Grison, P., Ferlet, R., Vidal-Madjar, A., Maurice, E., Pr\'{e}vot, L., et al., 1997, A\&A, 324, 471 
\bibitem[\protect\citeauthoryear{Sebo et al.}{2002}]{seb02} Sebo, K., Rawson, D., Mould, J., Madore, B., Putman, M., Graham, J., Freedman, W., Gibson, B. \& Germany, L., 2002, ApJS, 142, 71 
\bibitem[\protect\citeauthoryear{Shobbrook}{1992}]{sho92} Shobbrook, R., 1992, MNRAS, 255, 486
\bibitem[\protect\citeauthoryear{Simon \& Lee}{1981}]{sim81} Simon, N. \& Lee, A., 1981, ApJ,  248, 291
\bibitem[\protect\citeauthoryear{Simon \& Moffett}{1985}]{sim85} Simon, N. \& Moffett, T., 1985, PASP, 97, 1078
\bibitem[\protect\citeauthoryear{Simon et al.}{1993}]{sim93} Simon, N., Kanbur, S. \& Mihalas, D., 1993, ApJ, 414, 310
\bibitem[\protect\citeauthoryear{Skrutskie}{1998}]{skr98} Skrutskie, M., 1998, in {\it The Impact of Near-Infrared Sky Surveys on Galactic and Extra-galactic Astronomy}, eds. N. Epchtein, Astrophysics \& Space Science Library Vol. 230, Kluwer Academic Pub., Dordrecht, pg. 11
\bibitem[\protect\citeauthoryear{Storkey et al.}{2004}]{sto04} Storkey, A. J., Hambly, N. C., Williams, C. K. I. \& Mann, R. G., 2004, MNRAS, 347, 36
\bibitem[\protect\citeauthoryear{Tammann \& Reindl}{2002}]{tam02} Tammann, G. A. \& Reindl, B., 2002, Ap\&SS, 280, 165 
\bibitem[\protect\citeauthoryear{Tammann et al.}{2002}]{tam02a} Tammann, G. A., Reindl, B., Thim, F., Saha, A. \& Sandage, A., 2002, in {\it A New Era in Cosmology}, eds. N. Metcalfe \&  T. Shanks, ASP Conf. Series Vol. 283, Astron. Soc. Pac., San Francisco, pg. 258
\bibitem[\protect\citeauthoryear{Tanvir}{1997}]{tan97} Tanvir, N., 1997, in \textit{The Extra-galactic Distance Scale}, STScI May Symposium, eds. Livio, Donahue \& Panagia, Cambridge University Press, pg. 91
\bibitem[\protect\citeauthoryear{Thim et al.}{2003}]{thi03} Thim, F., Tammann, G., Saha., A., Dolphin, A., Sandage, A., Tolstoy, E. \& Labhardt, L., 2003, ApJ, 590, 256
\bibitem[\protect\citeauthoryear{Thim et al.}{2004}]{thi04} Thim, F., Hoessel, J., Saha, A., Claver, J., Dolphin, A. \& Tammann, G., 2004, AJ, 127, 2322
\bibitem[\protect\citeauthoryear{Udalski et al.}{1999a}]{uda99} Udalski, A., Szyma\'{n}ski, M., Kubiak, M., Pietrzy\'{n}ski, G., Soszy\'{n}ski, I., Wozniak, P. \& Zebrun, K, 1999a, AcA, 49, 201
\bibitem[\protect\citeauthoryear{Udalski et al.}{1999b}]{uda99b} Udalski, A., Soszy\'{n}ski, I., Szyma\'{n}ski, M., Kubiak, M., Pietrzy\'{n}ski, G., Wozniak, P. \& Zebrun, K., 1999b, AcA, 49, 223 
\bibitem[\protect\citeauthoryear{Weisberg}{1980}]{wei80} Weisberg, S., 1980, {\it Applied Linear Regression}, John Wiley \& Sons, $1^{st}$ Ed.
\bibitem[\protect\citeauthoryear{Zaritsky et al.}{2004}]{zar04} Zaritsky, D., Harris, J., Thompson, I. \& Grebel, E., 2004, AJ, 128, 1606
\end{thebibliography}
\end{document}